\documentclass[12pt]{article}

\usepackage{graphicx}
\usepackage{lscape}

\begin{document}

\title{Frictional sliding modes
along an interface between identical elastic plates
subject to shear impact loading}

\author{D. Coker$^a$, G. Lykotrafitis$^b$, A. Needleman$^a$ and 
A.J. Rosakis$^b$ \\
$^a$Division of Engineering, Brown University\\
Providence, RI 02912 \\
$^b$Division of Engineering and Applied Science\\
California Institute of Technology\\
Pasadena, CA 91125}

\maketitle

\begin{abstract}
\noindent Frictional sliding along an interface between two identical
isotropic elastic plates under impact shear loading is investigated
experimentally and numerically. The plates are held together by a
compressive stress and one plate is subject to edge impact near the
interface. The experiments exhibit both a crack-like and a pulse-like
mode of sliding. Plane stress finite element calculations modeling the
experimental configuration are carried out, with the interface
characterized by a 
rate and state dependent frictional law. 
For low values of the initial compressive stress
and impact velocity, sliding occurs in a crack-like mode. For higher values
of the initial compressive stress and/or impact velocity, sliding takes
place in a pulse-like mode. A variety of sliding modes are obtained in
the calculations depending on the impact velocity, the
initial compressive stress and the values of interface variables. 
One pulse-like mode involves well-separated
pulses with the pulse amplitude increasing with propagation distance.
Another pulse-like mode involves a pulse train of essentially constant
amplitude. The propagation speed of the leading pulse (or of the tip of the
crack-like sliding region) is near the longitudinal wave speed and never
less than $\sqrt{2}$ times the shear wave speed. Supersonic trailing pulses
are seen both experimentally and computationally. The trends in the
calculations are compared with those seen in the experiments.
\end{abstract}

\thispagestyle{empty}

\thispagestyle{empty}

\newpage

\section{Introduction}

Frictional sliding along an interface between two rapidly 
deforming solids is a
basic problem of mechanics that arises in a variety of contexts including
moving machinery surface interaction (both macro and micro
machines), material processing (e.g. cutting), the failure of fiber
reinforced composites (e.g. fiber pullout) and earthquake dynamics
(fault rupture). However, a framework for 
quantifying the wide range of observed dynamic frictional phenomena is only
beginning to emerge. The classical Amontons-Coulomb description of friction
states that the shear stress at an interface is proportional to the
normal stress, with the coefficient of proportionality being the
coefficient of friction. Two coefficients of
friction are identified; a static coefficient of friction that governs the
onset of sliding and a dynamic coefficient of friction that characterizes
the behavior during sliding.

At the microscale, an evolving population of deforming and fracturing
contacts, possible phase transitions and the presence of various
lubricants play an important role in setting the static and dynamic
coefficients of friction as well as in governing the transition between
them. Rate and state models of friction aim at incorporating the
effects of these microscale processes through appropriately chosen
state variables, e.g. Dieterich (1979), Ruina (1983), Rice and Ruina
(1983), Linker and Dieterich (1992), Prakash and Clifton (1993), and
Prakash (1998).  

Rate and state models of friction
have come to the fore because they substantially influence
the predicted mode and stability of sliding. Of particular interest is
whether sliding occurs in a crack-like 
mode in which the surfaces behind the leading edge of sliding
continuously slide or in a pulse-like mode, first proposed by
Heaton (1990), in which sliding occurs over a relatively small
propagating region. One significance of the sliding mode is that
the calculated frictional dissipation in the pulse-like mode is
significantly less than in the crack-like mode and is consistent with
some values of heat generation inferred from geophysical field
measurements (Heaton, 1990).

The classical Amontons-Coulomb description of friction is
inadequate for 
addressing fundamental issues of sliding along interfaces between
elastic solids because with Amontons-Coulomb friction sliding along such an
interface is unstable to
periodic perturbations for a wide range of friction coefficients and
material properties, with a growth rate proportional to the wave
number (Renardy, 1992; Adams, 1995). When generalized Rayleigh waves exist,
Ranjith and Rice (2001) found that there are unstable modes for all
values of the friction coefficient. Mathematically, instability of
periodic perturbations renders the response of a material interface with
Amontons-Coulomb friction ill-posed. Physically, it implies that
during sliding 
energy is transferred to shorter wave lengths, leading to pulse sharpening
and splitting. Numerically, the splitting of individual pulses creates an
inherent grid-size dependence (Andrews and Ben
Zion, 1997; Ben Zion and Andrews, 1998). Ranjith and Rice (2001)
showed that an experimentally based rate and state dependent friction
law (Prakash-Clifton, 1993; Prakash, 1998), in which the shear strength in
response to an abrupt change in normal stress evolves continuously with
time, regularizes the problem. Convergence through grid size reduction is
then achieved (Cochard and Rice, 2000). 

Although rate- and state-dependent frictional laws provide 
regularization, slip pulses may
still grow with time. Perrin et al. (1995) and Zheng and Rice (1998) found
from anti-plane strain (mode-III) calculations conditions on the frictional
law and the loading conditions needed for well-posedness. They also found
that under certain conditions, which include the rate of steady state
velocity weakening, sliding can be partially crack-like
and partially pulse-like. Ben-Zion and Huang (2002) numerically analyzed a
configuration of two identical solids separated by a variable-width fault zone
layer and assumed constant external compressive and shear fields. They
showed that the friction law used by Ranjith and Rice (2001) and Cochard and
Rice (2000) regularizes the Adams instability for identical as well
as for dissimilar
materials. The same conclusion was reached analytically by Rice et al.
(2001). The use of rate and state dependent friction not only regularizes
the ill-posedness of the sliding problem but can also eliminate supersonic
propagation (Rice et al., 2001). However, the physical reason for 
excluding supersonic propagation is not clear.

The issue of limiting propagation speeds arises in the dynamic fracture
mechanics of growing shear cracks (Freund, 1998; Broberg, 1999;
Rosakis, 2002) which has many similarities with the frictional
sliding process. Early
analytical and numerical studies of intersonic shear crack growth
using continuum slip weakening (Burridge, 1973; Andrews, 1976;
Broberg, 1995, 1996) and velocity weakening cohesive zone models
(Samudrala et al., 2002) concentrated on elucidating the mechanism of
transition from sub-Rayleigh to super-shear wave propagation speeds,
see also Madariaga and Olsen (2000) and Dunham et al. (2003), and on
the identification of stable intersonic propagation speed regimes.  Of
particular relevance to the present study is the persistent occurrence
of intersonic shear rupture along interfaces between identical solids
which was first observed experimentally by Rosakis et al. (1999) and
by Coker and Rosakis (2001). This phenomenon was subsequently modeled
numerically by Needleman (1999) and by Hao et al. (2004). Atomistic
models of intersonic shear rupture (Abraham and Gao, 2000; Abraham,
2001; Gao et al., 2001), field observations of intersonic rupture
events during recent large crustal earthquakes (Archuleta, 1984; Olsen
et al., 1997; Hernandez et al., 1999; Bouchon et al., 2001; Lin et
al., 2002; Bouchon and Valle, 2003), as well
as recent laboratory models of earthquake rupture (Xia et al., 2004)
have demonstrated the remarkable length scale persistence (over eleven 
orders of magnitude) of intersonic rupture phenomena and of the main
features observed in laboratory experiments and in continuum theories.

Here, frictional sliding between identical rectangular plates subject
to an 
initial compressive stress and impact shear loading is studied both
experimentally and numerically. In the experiments, a
uniform compressive stress is applied to two Homalite plates and the impact
loading is imposed using a gas gun and a steel projectile. Dynamic
photoelasticity is used to record fringe patterns on a micro-second
time-scale. The fringe patterns give indirect evidence of the sliding
mode and the propagation speed of the sliding tip is measured. Plane
stress calculations modeling the experimental configuration are
carried out with friction 
characterized by a rate- and state-dependent relation, including the
Prakash-Clifton (Prakash and Clifton, 1993; Prakash, 1998) normal
stress dependence. 
A parameter study is undertaken to predict possible modes of sliding
in this configuration together with the associated propagation speeds.
The dependence on loading conditions is explored and the
computational results are compared with the experimental
observations. 

\section{Formulation}

\subsection{Boundary value problem}

We consider two solids undergoing relative motion across an interface 
$S_{int}$. A framework is used where 
two constitutive relations are specified; one for the
plate material and one to characterize frictional sliding in terms of
the traction and velocity jump across the interface. 
Plane stress conditions are assumed, geometry 
changes are neglected\footnote{The calculations are carried out 
using a finite element code developed for 
finite deformations in which the constitutive relation is expressed in terms
of second Piola-Kirchhoff stress and Lagrangian strain components; otherwise
finite deformation effects are neglected in the calculations here.} and the
principle of virtual work is written as 
\begin{equation}
{\int_{A}}\mbox{\boldmath $\sigma$}:
{\delta \mbox{\boldmath $\epsilon$}}{dA}-
{\int_{S_{int}}}\mathbf{T}\cdot {\delta \mathbf{\Delta u} {dS}}=
{\int_{S_{ext}}}\mathbf{T}\cdot {\delta \mathbf{u}{dS}}-{\int_{A}}\rho 
{\frac{{\partial ^{2}\mathbf{u}}}{{\partial t^{2}}}}\cdot 
{\delta \mathbf{u}{dA}}
\label{eqfe2}
\end{equation}
\noindent where $t$ is time, $\mbox{\boldmath $\sigma$}$ is the stress
tensor, $\mbox{\boldmath $\epsilon$}$ is the strain tensor, $\mathbf{u}$ is
the displacement vector, $\mathbf{T}$ is the traction vector, 
$\mathbf{\Delta u}$ is the displacement jump across the interface,
$\mbox{\boldmath $\sigma$}:
\mbox{\boldmath $\epsilon$}$ denotes $\sigma_{ij}\epsilon_{ji}$, and 
$A$, ${S_{ext}}$ and ${S_{int}}$ are 
the area, external boundary, and interface line, respectively, in the
reference configuration. 

\begin{figure}[tbp]
\begin{center}
\includegraphics[width=3.7in]{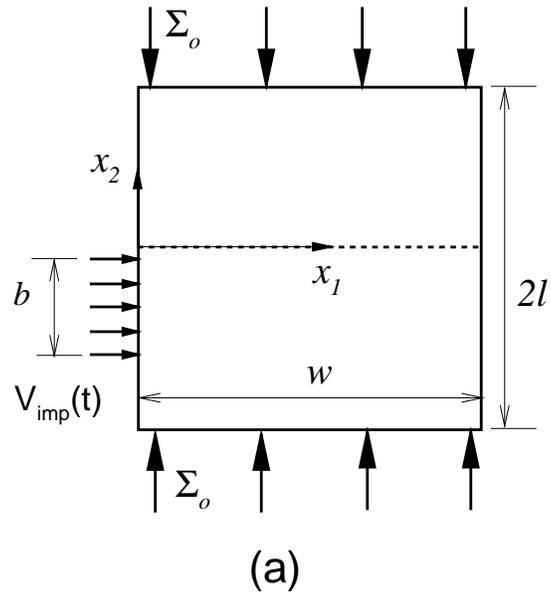}\\[0pt]
\includegraphics[width=3.2in]{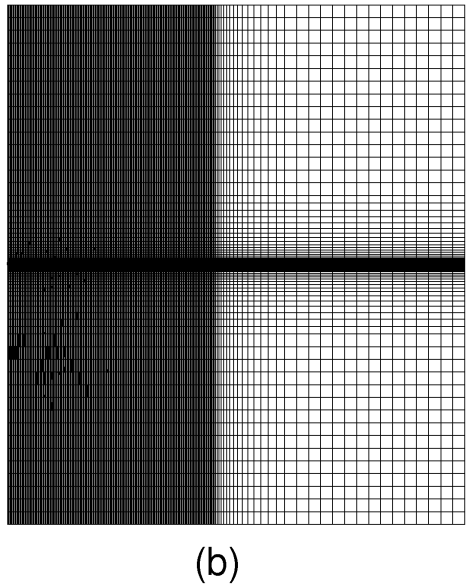}
\end{center}
\caption{(a) Geometry and loading configuration used in the
plane stress finite element calculations. (b) The finite element mesh
used in the calculations ($56,580$ degrees of freedom).}
\label{fig1}
\end{figure}

Computations are carried out for the specimen geometry shown in
Fig.~\ref{fig1} with $\ell=75$ mm and $w=137$ mm. A Cartesian
coordinate system with 
the $x_1-x_2-$plane being the plane of deformation is used and the origin is
taken as shown in Fig.~\ref{fig1}. The plates are subject to a uniform
compressive stress of magnitude $\Sigma_0$. For $t \le 0$, $S_{22}=-\Sigma_0$
and $T_n=\Sigma_0$ along the interface. Displacements are measured from the
uniformly compressed state at $t=0^-$.

At $t=0^+$, a normal velocity is prescribed along the portion of the
edge $x_1=0$ for which $-b\leq x_2 \leq 0$, with $b=25$ mm, and the
shear traction 
is taken to vanish there. Hence, 
\begin{equation}
u_{1}=- \int_0^t V(\xi) d \xi \quad ,\quad T_{2}=0\qquad \hbox {on $x_1=0$
and $-b \le x_2 \le 0$}  \label{eqbc}
\end{equation}
where in eq.~(\ref{eqbc}) 
\begin{equation}
V(t)=\left\{ 
\begin{array}{cccc}
V_{\mathrm{imp}}\,t/t_{r}, & \hbox {for $0 \le t < t_r$;} &  &  \\ 
V_{\mathrm{imp}}, & \hbox {for $ t_r \le t \le t_p$} &  &  \\ 
V_{\mathrm{imp}}\,[1-(t-t_p)/t_s], & \hbox {for $ t_p< t <(t_p+ t_s)$} &  & 
\\ 
0, & \hbox {for $ t \ge (t_p+t_s)$} &  & 
\end{array}
\right.  \label{vpulse}
\end{equation}
Here, $t_r$ is the rise time, $t_p$ is the pulse time and $t_s$ is the step
down time. In the calculations $t_r$ and $t_s$ are fixed 
at $10$ $\mu$s and $(t_p-t_r)$ is $50$ $\mu$s. On the remaining
external surfaces of the 
specimen $\mathbf{T}=\mathbf{0}$.

The plate material is characterized as an isotropic
elastic solid with Young's modulus $E$ and Poisson's ratio $\nu$, with
properties representative of
Homalite: $E=5.2$ GPa, $\nu=0.34$ and density $\rho=1230$ kg/m$^3$. These
give elastic wave speeds of
\begin{equation}
c_l=2187 \, \rm{m/s} \quad c_s=1255  \, \rm{m/s} \quad c_R=1155 \,
\rm{m/s} 
\label{eq:R}
\end{equation}
where $c_l$ is the longitudinal wave speed, $c_s$ is the shear wave
speed and $c_R$ is the Rayleigh wave speed.

A rate- and state-dependent friction law is taken to characterize the
response along the interface $x_2=0$. With $n$ and $s$, respectively,
denoting in-plane directions normal and tangential to the interface, 
$\Delta \dot{u}_n$
and $\Delta \dot{u}_s$ are the difference in velocity between material
points that are on opposite sides of the interface in the initial
configuration. Since the change in the positions of material points
along the interface is not accounted for, the formulation is
restricted to small amounts of sliding. When 
$\Delta {u}_n > -\Sigma_0/C_n$, the interface is in tension and 
\begin{equation}
T_n=0 \qquad T_s=0
\end{equation}
along $S_{int}$. Otherwise, 
\begin{eqnarray}  \label{eqtime11}
\dot{T}_n &=& -C_n \Delta \dot{u}_n \\
\dot{T}_s &=& C_s [\Delta \dot{u}_s-\mbox{{\rm sgn}}{(T_s)} 
\Delta \dot{u}_{\mathrm{slip}}].  
\label{eqtime1}
\end{eqnarray}
Here, as in Povirk and Needleman (1993), the interface is presumed to have
an elastic stiffness and $\Delta \dot{u}_{\mathrm{slip}}$ is the magnitude
of the sliding velocity specified by the rate- and state-dependent friction
law ($\Delta \dot{u}_{\mathrm{slip}} \ge 0$). In the elastic-friction
relation eq.~(\ref{eqtime1}) $\Delta \dot{u}_{\mathrm{slip}}$ is an
internal variable determined from the frictional constitutive
description, while $\Delta \dot{u}_s$ is the jump in
particle velocity across the interface. Unless otherwise specified,
the values $C_n=0.3$ MPa/m and  
$C_s=0.1$ MPa/m are used in the calculations here. 

The finite element discretization is based on linear displacement triangular
elements that are arranged in a `crossed-triangle' quadrilateral pattern.
When the finite element discretization of the displacement field is
substituted into eq. (\ref{eqfe2}) and the integrations are carried out,
equations of the form 
\begin{equation}
\mathbf{M}\frac{\partial^2 \mathbf{U} }{\partial t^2}=\mathbf{R}
\label{eqfe4}
\end{equation}
are obtained where $\mathbf{U}$ is the vector of nodal 
displacements, $\mathbf{M}$ is the mass matrix and $\mathbf{R}$ is the
nodal force vector 
consisting of contributions from the area elements and the interface. 
Four point Gaussian integration is used along the
interface. These finite element equations are integrated numerically
by an explicit integration procedure, the Newmark $\beta$-method
(Belytschko et al., 1976). A lumped mass matrix is used instead of the
consistent mass matrix, since this has been found preferable for
explicit time integration procedures, from the point of view of
accuracy as well as computational efficiency (Krieg and Key, 1973).

The mesh used in the calculations consists of a fine region with uniform
rectangular elements near the interface at impact edge, with the mesh
spacing then gradually increasing in size. The mesh has $18,320$
quadrilateral elements and $56,580$ degrees of freedom, with the uniform
mesh region consisting of $200 \times 20$ rectangles, each of which is $0.30$
mm $\times$~$0.20$ mm, so that the uniform mesh extends $60$ mm from the
impact edge. The configuration analyzed and the finite element mesh used are
shown in Fig.~\ref{fig1}.

\subsection{Rate- and state-dependent friction law}

The classical Amontons-Coulomb frictional relation has the form 
\begin{equation}
\vert {T}_s \vert = \mu {T}_n  \label{coul1}
\end{equation}
where ${T}_n$ is the component of ${\mathbf{T}}$ in the direction of the
normal to the interface, ${T}_s$ is the component of ${\mathbf{T}}$ tangent
to the sliding direction and $\mu$ is the (constant) coefficient of
friction. The relation (\ref{coul1}) holds when the normal traction is
compressive, i.~e. with the sign convention here when ${T}_n>0$.

The problem of two elastic half-spaces sliding with a constant
coefficient of friction, $\mu $, is ill-posed for a significant range
of values of $\mu $, Adams (1995). This manifests itself in
numerical solutions through a lack of convergence with increasing mesh
refinement.
Rate and state dependent friction laws were introduced, mainly in the
geophysics literature, to account for experimental observations that could
not be rationalized in the context of a Coulomb friction description. At
each point on the bounding surface, the coefficient of friction $\mu$ in eq.
(\ref{coul1}) is taken to depend on the frictional sliding velocity at that
point, $\Delta \dot u_{\mathrm{slip}}$, and a 
set of state variables, $\theta_i$ so that 
\[
\vert T_s \vert =\mu(\theta_i, \Delta \dot{u}_{\mathrm{slip}}) T_n 
\]

The state variables are regarded as phenomenological parameters that account
for the change of contact quality between the surfaces over time. Rate and
state dependent friction models can account for the following fundamental
observations on friction (Dieterich, 1979; Rice and Ruina, 1983; Ruina,
1983):

\begin{itemize}
\item[(i)]  there is an instantaneous increase in the coefficient of
friction in response to a step increase $\Delta \dot
u_{\mathrm{slip}}$ (the direct effect),

\item[(ii)]  there is a subsequent change to a steady-state value of
the coefficient of friction with this steady-state value being a decreasing
function of $\Delta \dot u_{\mathrm{slip}}$, and

\item[(iii)]  the approach to this steady-state value occurs over a
characteristic distance that is independent of $\Delta \dot
u_{\mathrm{slip}}$.
\end{itemize}

Another key observation is that transmitted shear stresses do not
instantaneously follow a step drop in the normal stresses, Prakash and
Clifton (1993), Prakash(1998). Instead, immediately following a step drop in 
$\vert T_n \vert$, as the slip accumulates the shear stress gradually
approaches a new steady-state level characteristic of the new normal
pressure and the current sliding velocity, Prakash and Clifton (1993). The
relation, eq. (\ref{coul1}), is replaced by, Prakash and Clifton (1993),
Prakash(1998), 
\begin{equation}
\vert {T}_s \vert =\mu(\theta_0,\Delta \dot{u}_{\mathrm{slip}})
(\theta_1+\theta_2) 
\label{pc1}
\end{equation}
with the evolution of the state variables $\theta_1$ and $\theta_2$ given by 
\begin{eqnarray}
\dot{\theta}_1 &=& -\frac{1}{L_1} [ \theta_1 - C T_n]
\Delta \dot{u}_{\mathrm{slip}}, \\
\dot{\theta}_2 &=& -\frac{1}{L_2} [ \theta_2 - D T_n]
\Delta \dot{u}_{\mathrm{slip}}.  \label{pc2}
\end{eqnarray}
where $L_1$ and $L_2$ are characteristic lengths. In the steady 
state, when $\dot{\theta}_1=0$ and $\dot{\theta}_2=0$,
eqs. (\ref{pc1}) to (\ref{pc2}) imply $T_s \propto T_n$.

In Dietrich (1979), Rice and Ruina (1983) and Ruina (1983) $\mu$ was taken
to have a logarithmic dependence on the sliding velocity. In such a
relation, $\mu$ is not bounded at $\Delta \dot{u}_{\mathrm{slip}}=0$ which
leads to difficulties in numerical computations. Rice and Ben-Zion (1996)
and Ben-Zion and Rice (1997) appealed to an activated rate process
interpretation which resulted in a relation bounded at 
$\Delta \dot{u}_{\mathrm{slip}}=0$. Here, we adopt a purely
phenomenological expression for $\mu$ proposed by Povirk and Needleman
(1993) that has the form  
\begin{equation}
\mu(\theta_0, \Delta \dot{u}_{\mathrm{slip}})=g(\theta_0) 
\Biggr ( \frac{\Delta \dot{u}_{\mathrm{slip}}} {V_0} + 1 \Biggl
)^{1/m}  
\label{pn1} 
\end{equation}
with 
\begin{equation}
g(\theta_0) =\frac {\mu_d+(\mu_s-\mu_d)\exp \Biggr[ -\Biggr (\displaystyle{\ 
\frac{L_0/\theta_0}{V_1}} \Biggl )^p \Biggl ] } 
{\Biggr[ \displaystyle{\frac{L_0/\theta_0}{V_0} + 1 \Biggl ]^{1/m}}},
\label{pn2} 
\end{equation}
where the evolution of $\theta_0$ is given by, 
\begin{equation}
\dot{\theta}_0=B \Biggl ( 1- \frac{\theta_0 \Delta 
\dot{u}_{\mathrm{slip}}}{L_0} \Biggr ) 
\label{eqthetav}
\end{equation}
and $\mu_s$, $\mu_d$, $L_0$, $V_0$, $V_1$ and $B$ are prescribed constants.

The steady state value of $\theta_0$, i.e. the value at which $\dot\theta_0=0
$ is $L_0/\Delta \dot{u}_{\mathrm{slip}}$. Substituting this value into eqs.
(\ref{pn1}) and (\ref{pn2}) gives the steady state value for the friction
coefficient, $\mu_{ss}$, at sliding velocity $\Delta \dot{u}_{\mathrm{slip}}$
as 
\begin{equation}
\mu_{ss}= \mu_d+(\mu_s-\mu_d)\exp \Biggr[ -\Biggr 
(\displaystyle{\ \frac{\Delta \dot{u}_{\mathrm{slip}}}{V_1}} 
\Biggl )^p \Biggl ]  
\label{eqss}
\end{equation}

The frictional relation eq.~(\ref{eqtime1}) is only 
invoked if there is 
contact ($T_n \ge 0$). Even when there is contact, 
$\Delta \dot{u}_{\mathrm{slip}}$ may still vanish. Define 
\begin{equation}
\beta =\frac{|T_s|} {g(\theta_0) (\theta_1+\theta_2)},
\label{eqbeta}
\end{equation}
then from eqs.~(\ref{pc1}) and (\ref{pn1}), and requiring $\Delta
\dot{u}_{\mathrm{slip}}$ to be 
non-negative, $\Delta \dot{u}_{\mathrm{slip}}$ is given by 
\begin{equation}
\Delta \dot{u}_{\rm slip}= \left\{ \begin{array}{ll}
V_0( \beta^m -1 ) & \textrm{ for $\beta>1$}  \\
 0                      & \textrm{ for $\beta \leq 1$ }
\end{array} \right.
\label{eqtime3}
\end{equation}
Note that $\beta >1$ implies $|T_s|-g(\theta_0)(\theta_1+\theta_2)>0$. When
the normal traction is constant $\theta_1+ \theta_2= T_n$ and $\beta$
simplifies to $|T_s|/T_n g(\theta_0)$ as in Povirk and Needleman (1993).

We refer to circumstances where $\Delta \dot{u}_{\mathrm{slip}}=0$ as
sticking and circumstances where $\Delta \dot{u}_{\mathrm{slip}}
>0$ as frictional sliding.

The procedure used to carry out time integration of this frictional
constitutive relation is described in the Appendix.

\subsection{Friction law parameters}

Samudrala et al. (2002) and Rosakis (2002) fit sliding experiments of
Homalite on Homalite (Rosakis et al., 1999; 2000) with an expression of the
form 
\begin{equation}
\mu_{ss}=\mu_0 \left [ 1+\alpha \frac{\Delta \dot{u}_{\mathrm{slip}}}{c_s} 
\frac{G}{2\tau_0} \right ]  \label{sr1}
\end{equation}
where $2G=E/(1+\nu)$ and $\tau_0$ is the shear strength of Homalite. From
the observed inclination of tensile micro-cracks emanating from the
interface, the parameter values $\alpha=-0.4$, $G/\tau_0=136$, and $\mu_0=0.6$
were obtained.

\begin{figure}[tbp]
\begin{center}
\includegraphics[width=5in]{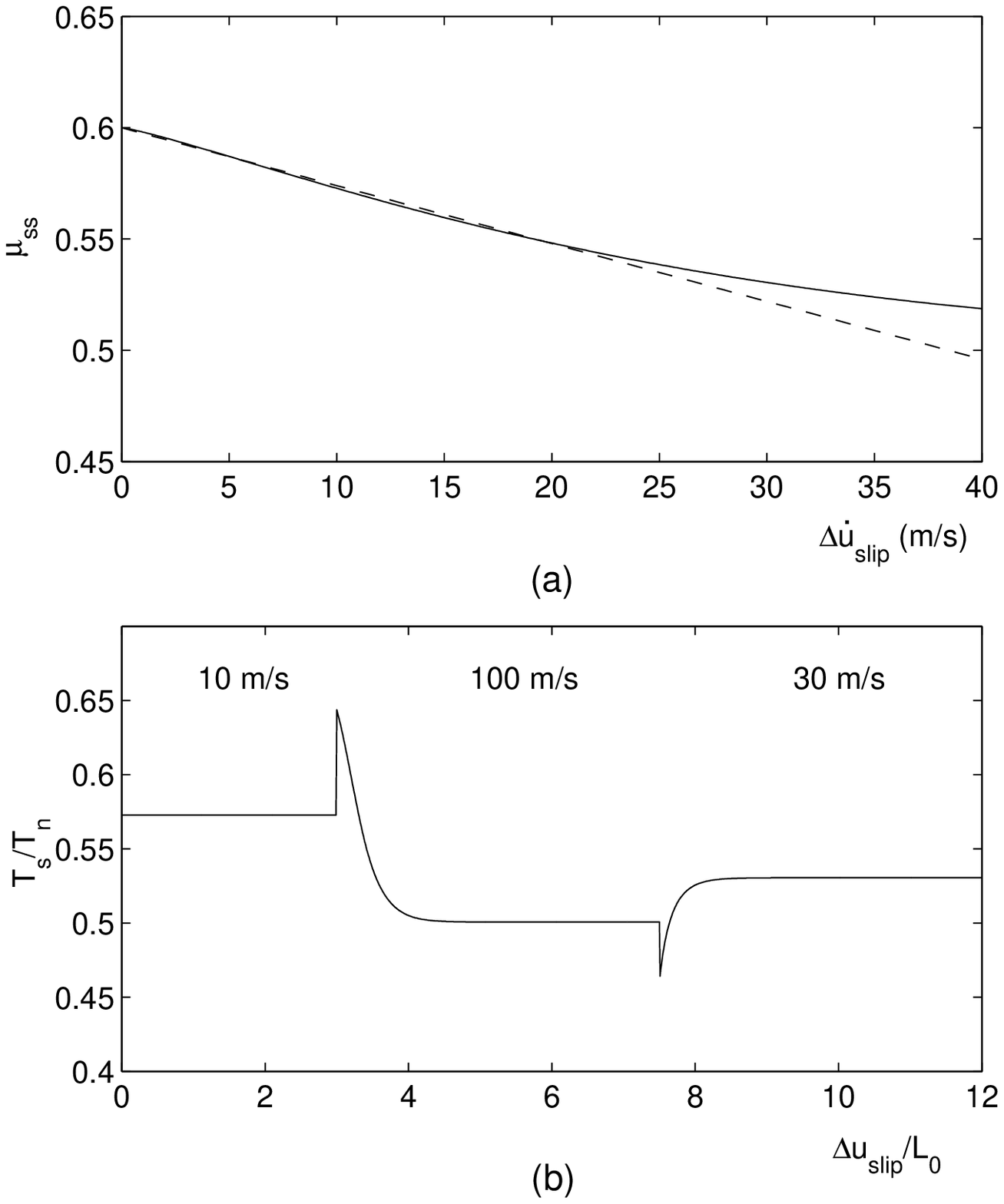}
\end{center}
\caption{(a) Comparison of the steady-state coefficient of friction, 
$\mu_{ss}$, as a function of $\Delta \dot{u}_{\mathrm{slip}}$ for 
eq.~(\ref{sr1}) (dashed line) and for eq.~(\ref
{eqss}) using the properties in Table~\ref{tab:3} (solid line). (b) The
effect of an abrupt change in $\Delta \dot{u}_{\mathrm{slip}}$ on the apparent
coefficient of friction $T_s/T_n$ for the rate- and state-dependent friction
relation used in the calculations.}
\label{fig2}
\end{figure}

Parameter values in eq.~(\ref{eqss}) were chosen to match the response in
eq.~(\ref{sr1}) for $\Delta \dot{u}_{\mathrm{slip}}<40$ m/s. Figure \ref
{fig2}a shows $\mu_{ss}$ as a function of $\Delta \dot{u}_{\mathrm{slip}}$
from eq.~(\ref{sr1}) and from eq.~(\ref{eqss}) with 
$\mu_s=0.6$, $\mu_d=0.5$, $p=1.2$ and $V_1=26$ m/s. The parameters
$V_0$ and $m$ in eq.~(\ref{pn1}) 
were chosen to be consistent with the observed 
response for $\Delta \dot{u}_{\mathrm{slip}}$ ranging from 10 m/s to
40 m/s (Fig.~\ref{fig2}a).  Figure \ref{fig2}b shows the 
friction coefficient, $T_s/T_n$, as a function of 
of the 
accumulated frictional sliding $\Delta u_{\mathrm{slip}}=\int
\Delta \dot u_{\mathrm{slip}}dt$. There is
an instantaneous increase in $T_s/T_n$
in the same direction as the change in $\Delta
\dot u_{\mathrm{slip}}$ 
(the direct effect) followed by a gradual decrease to a steady-state
value. The apparent coefficient of friction $T_s/T_n$ can attain values
outside the range defined by the static and dynamic 
coefficients of friction due to the direct effect in the rate- and
state-dependent friction law.

\begin{figure}[tbp]
\begin{center}
\includegraphics[width=5in]{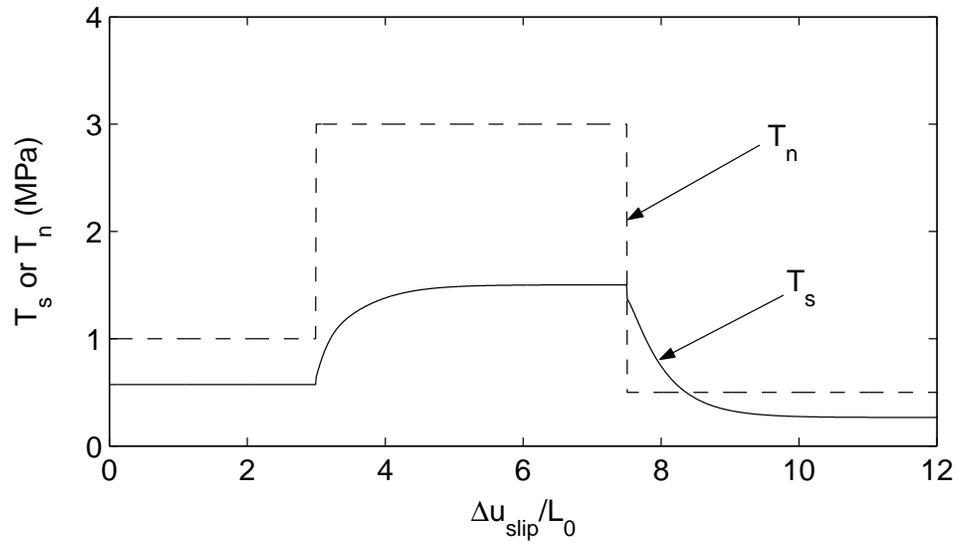}
\end{center}
\caption{Variation of the shear traction, $T_s$, (solid
line) as a function of $\Delta u_{\mathrm{slip}}$ for a step jump in
compressive normal traction, $T_n$, (dashed line) at constant frictional
sliding rate $\Delta \dot{u}_{\mathrm{slip}}$ for the rate- and
state-dependent friction relation used in the calculations.}
\label{fig3}
\end{figure}

The parameter values for the normal stress dependent response, 
$C$, $D$, $L_1$, are chosen close to values obtained in the
experiments of Prakash and 
Clifton (1993). The initial values for the internal 
variables were set to $\theta_0(0)=L/10\dot{V}_1$ and $\theta_1(0) = C
\Sigma_0$ and $\theta_2(0) = 
D \Sigma_0$. Figure \ref{fig3} shows the response using
eqs. (\ref{pc1}) to (\ref{eqthetav}) 
under a step jump in compressive normal traction at a constant sliding
velocity. The plots show the imposed normal traction (dashed line)
and the shear traction (solid line) as functions of the 
accumulated frictional sliding $\Delta u_{\mathrm{slip}}$.
There is a gradual change of 
the state variables and the shear resistance after a step decrease in the
normal compressive stress. The apparent coefficient of friction, $T_s/T_n$,
however, has a sudden jump.

The choice of the initial values for the state variables is arbitrary but
can be of significance when comparing to other friction models such as
slip-weakening models (Bizzarri et al.~2001). The parameters characterizing
the frictional constitutive relation used in the calculations are those in
Table~\ref{tab:3} unless specifically stated otherwise.

\section{Experimental Methods and Results}

The experimental procedures are similar to those used to study shear crack
propagation (Rosakis et al., 1999). Two Homalite plates, each with 
$\ell=139.7$ mm $w=76.2$ mm (see Fig.~\ref{fig1}a) and $9.525$ mm
thick are 
held together by a uniform compressive stress. Homalite is a brittle
polyester resin that exhibits stress induced birefringence and is mildly
rate sensitive. At a strain rate of $10^3$ s$^{-1}$ the elastic wave
speeds are given by eq.~(\ref{eq:R}). The impact 
loading is imposed via a cylindrical steel 
projectile of diameter 25 mm and length 51 mm fired using a gas gun with
impact speeds ranging from 10 m/s to 60 m/s. A steel buffer plate, $25.4$ mm 
$\times$ $76.2$ mm and $9.525$ mm thick, is bonded to the impacted plate to
prevent shattering at the impact side ($x_1=0$ in Fig.~\ref{fig1}a) and to
induce a more or less planar loading wave. A uniform compressive stress is
applied by a press which was calibrated using a load cell. The loading wave,
as measured from a strain gage glued to the specimen, is of a trapezoidal
form with a rise time of 10-20 $\mu$s followed by an essentially steady
velocity for 40 $\mu$s. Dynamic photoelasticity is used to extract stress
field information around the interface. The photoelastic fringe patterns
were recorded in real time using a high-speed Cordin CCD camera capable of
capturing 16 images at a rate of 100 million frames per second. A collimated
laser beam with a diameter of $130$ mm is used to illuminate the
specimen. Two pairs of circular 
polarizer plates, placed on either side of the specimen as described by
Rosakis et al. (1999), produce isochromatic fringes. The photoelastic
optical setup is arranged for light field. 

The isochromatic fringes are
related to contours of $\sigma_1-\sigma_2$, with $\sigma_1$ being the
maximum in-plane principal stress and $\sigma_2$ the minimum in-plane
principal stress, through the stress optical relation 
\begin{equation}
\sigma_1-\sigma_2=\frac{N F_{\sigma}}{h} ,
\end{equation}
where $F_{\sigma}$ is the stress optical coefficient of Homalite, $h$ is
the specimen thickness, and $N$ is the isochromatic fringe order.

\subsection{Experimental results}

The experimental results are presented in detail in Lykotrafitis et
al. (2004). Here, a few experimental results that illustrate
characteristic features are presented for comparison with the
numerical calculations.

\begin{figure}[tbp]
\begin{center}
\includegraphics[width=3.9in]{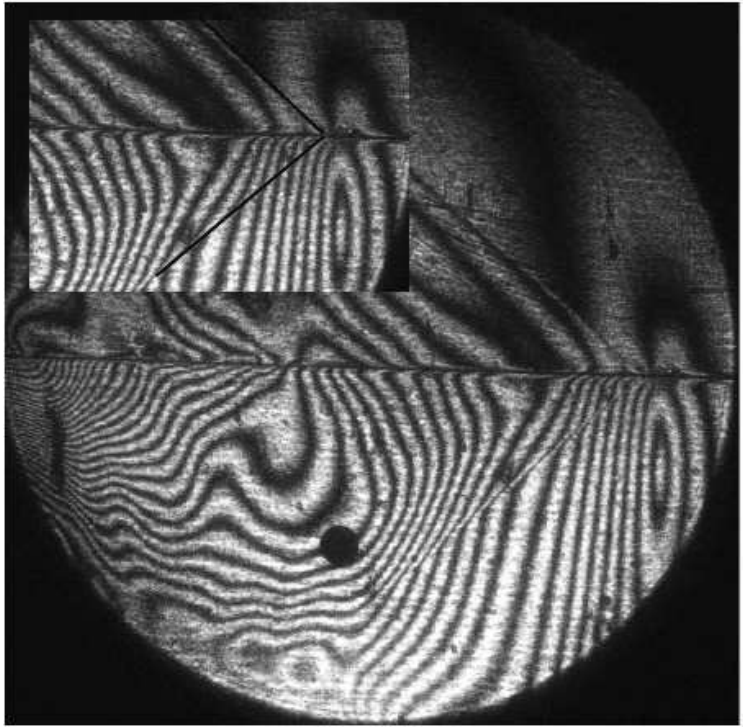}
\end{center}
\caption{Experimental isochromatic fringe patterns
from a dynamic
friction experiment on Homalite subject to a static compressive stress of
9.4 MPa and an impact velocity of 32.7 m/s. In the inset lines are
drawn to highlight Mach lines. The field of view is 130 mm in diameter.}
\label{figs4}
\end{figure}

Results with a compressive stress of 9.4 MPa and impact velocities of 32.7
m/s and 42.2 m/s are presented. Figure \ref{figs4} shows the
isochromatic fringe patterns obtained with an 
impact velocity of 32.7 m/s. Experiments
conducted at lower impact velocities exhibit similar characteristics. In
Fig.~\ref{figs4}, the loading wave front, which travels at $c_l$,
arrives from the left. Behind the loading wave front a shear
Mach cone, formed by a sharp change in fringe density is observed. The
inset to Fig.~\ref{figs4} focuses on the region 
near the propagating tip with a thick line drawn over the Mach line. The
sliding tip follows the loading wave and the tip can be located by
tracing the Mach line to the interface. In this case the sliding tip
follows shortly behind the dilational loading wave. The
propagation speed of the sliding tip is $\approx 1810$ m/s as
measured by two methods. 
One measurement involves following the position of the sliding tip in
various 
frames and knowing the frame timing, the propagation speed is
obtained. In the second method the Mach angle is used 
to obtain the tip speed. The two methods give consistent propagation
speeds. 

In Fig.~\ref{figs4}, there is a 
concentration of isochromatic fringes some distance behind the sliding
tip that propagates at approximately $c_R$  and that
is a consequence of loading wave reflections. Other
characteristic features in Fig.~\ref{figs4} are: (i) there is a cusp in
the stress contours at 
the interfaces, indicating that the propagation speed is slightly faster
along the interface than in the bulk; (ii) the fringe density is higher in
the plate where the impact loading is applied, showing that energy is not
transferred easily through the interface; and (iii) the fringe
discontinuity at the interface shows that sliding is occurring in
a crack-like mode. 

\begin{figure}
\begin{center}
\includegraphics[height=2.7in]{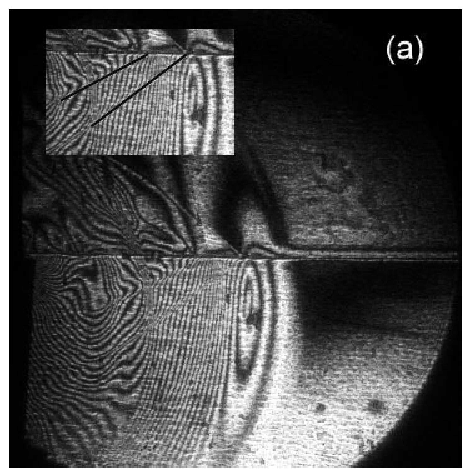}
\includegraphics[height=2.7in]{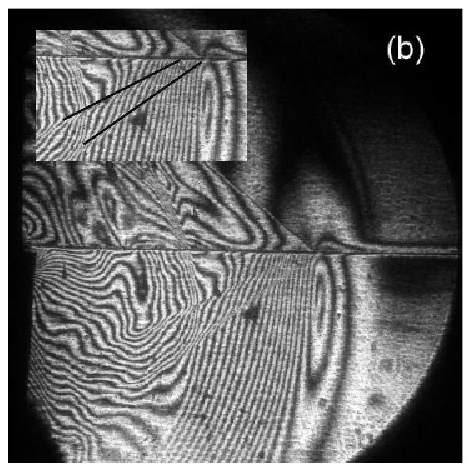}
\includegraphics[height=2.7in]{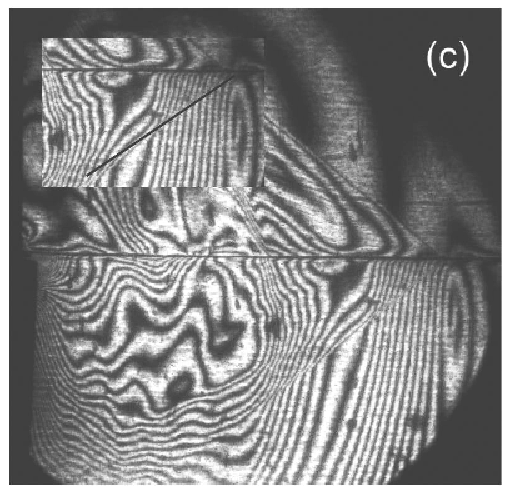}
\end{center}
\caption{Experimental isochromatic fringe patterns from a
dynamic friction experiment on Homalite subject to a static compressive
stress of 9.4 MPa and impact velocity of 42 m/s, at (a) $t=40 \mu s$, 
(b)$t=48 \mu s$, (c)$t=60 \mu s$. In the inset a line or lines are
drawn to highlight Mach lines. The field of view is 130 mm in diameter.}
\label{figs5}
\end{figure}

Isochromatic fringe patterns for an experiment with 
an impact velocity of 42 m/s are shown in Fig.~\ref{figs5} at
40 $\mu$s, 48 $\mu$s and 60 $\mu$s after impact. The thick lines in
the insets are drawn to show the Mach line locations and orientations.
Both methods of calculating the propagation speed of
the sliding tip give a propagation speed of about 1950 m/s.
Although the general characteristics
seen in Fig.~\ref{figs4} are preserved, new features
enrich the picture. Behind the impact wave a shear Mach line emanating from
the sliding tip is observed. In addition, a second Mach line emanates behind
the sliding tip that is not parallel to the first one
(Fig.~\ref{figs5}a). 
This Mach line is at a shallower slope corresponding to a supersonic
propagation speed of $\approx 2600$ m/s. In Fig.~\ref{figs5}b the tip
of the second 
Mach line approaches the tip of the first Mach line at the interface.
These two points merge as the second point catches up with the first
point (Fig.~\ref{figs5}c) and only one Mach line continues to be observed in
subsequent frames (not shown). 

Behind the second
Mach line there is a fringe concentration, associated
with the loading, that travels at the Rayleigh wave
speed. Some fringes pass continuously from the upper plate to the lower plate
which is consistent with a contact region forming.
In experiments with the magnitude of the initial compressive stress
increased, there is an
increase in the number of discontinuous regions of fringe
concentration that may indicate separate regions of sliding and
regions of sticking (i.e. multiple pulses). 
In each experiment, the speed of the leading sliding tip is found to be
essentially constant. In addition, the propagation speed of the
leading sliding tip
was found to increase with increasing impact speed (or decreasing
compressive 
stress) and in all cases to be between $\sqrt 2 c_s$ and $c_l$.

\section{Numerical Results}

The focus here is on illustrating the range of behaviors that are
obtained by varying the magnitude of the initial compressive stress $\Sigma_0
$ and the impact velocity $V_{\mathrm{imp}}$. Results are presented for
combinations of $\Sigma_0$ and $V_{\mathrm{imp}}$ that give
rise to the following five sliding modes:

\begin{description}
\item  {Case I:} A crack-like mode -- $\Sigma_0=6$ MPa,
  $V_{\mathrm{imp}}=2$ m/s. 

\item  {Case II:} A pulse-like mode -- $\Sigma_0=30$ MPa,
  $V_{\mathrm{imp}}=2$ m/s. 

\item  {Case III} A train of pulses -- $\Sigma_0=10$ MPa,
  $V_{\mathrm{imp}}=20$ m/s. 

\item  {Case IV:} Multiple pulses coalescing to form a crack --
  $\Sigma_0=0.9$ MPa, $V_{\mathrm{imp}}=10$ m/s. 

\item  {Case V:} A pulse-like mode followed by a crack-like mode --
  $\Sigma_0=40$ MPa, $V_{\mathrm{imp}}=2$ m/s. 

\end{description}

In all cases but Case V, the friction parameters are as 
specified in Table~\ref{tab:3}. For Case V, the exponent $p$ in
eq.~(\ref{pn2}) is taken to be $0.5$ and $V_1=1$ m/s. Also, for Case
IV the values of the interface elastic 
constants are $C_n=0.03$ MPa/m and $C_s=0.01$ MPa/m, one tenth their value
in all the other cases. In the results presented here, the frictional
sliding region remains within the fine uniform part of the mesh.

Results are presented for: (i) propagation speeds; (ii) the spatial
distribution and time evolution of various interface quantities; and
(iii) isochromatic fringe patterns (contours of $\sigma_1 -\sigma_2$
with $\sigma_1$ being the maximum in-plane principal stress and
$\sigma_2$ the minimum in-plane principal stress). 

\begin{figure}[tbp]
\begin{center}
\includegraphics[width=6.0in]{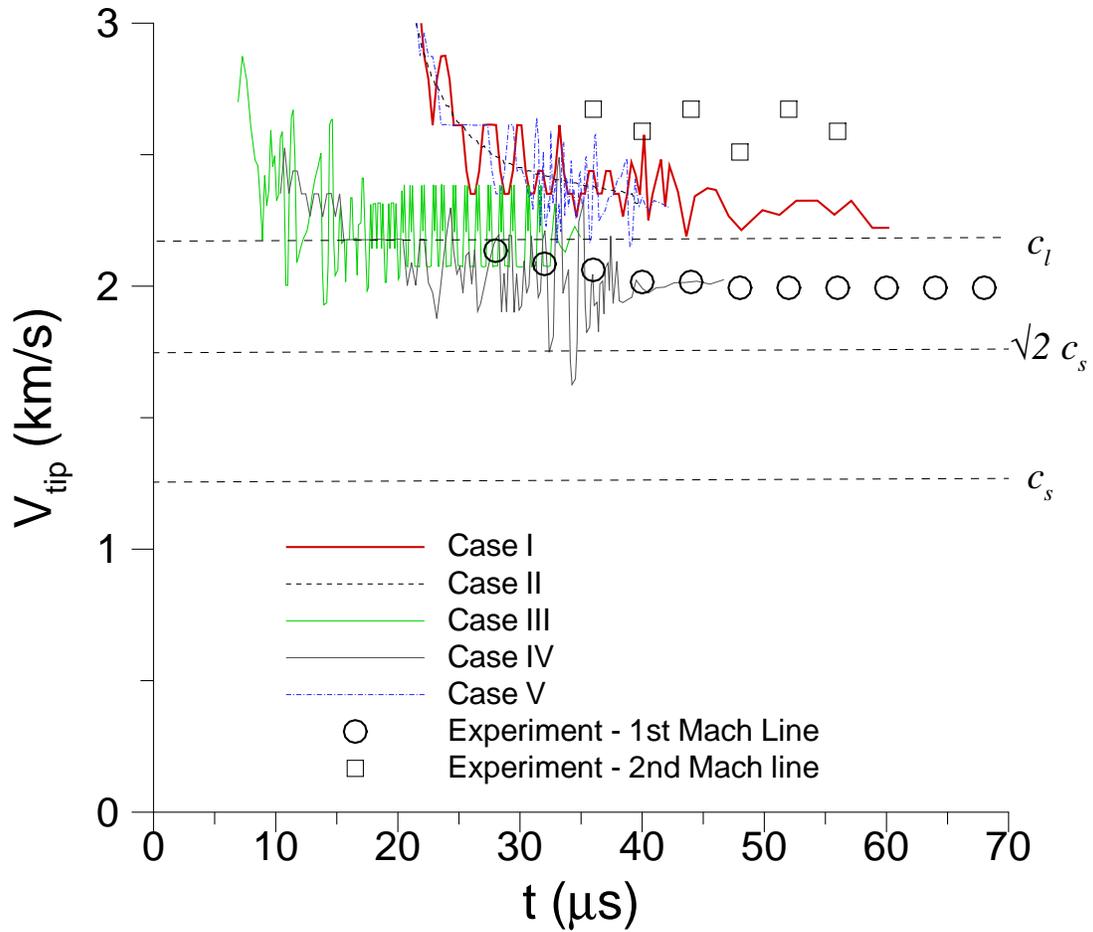}
\end{center}
\caption{Propagation speeds, $V_{\mathrm{rupt}}$, versus
time for the five cases analyzed. The symbols are the experimentally
measured propagation speeds from the Mach cone angle with $\Sigma_0=9.4$ MPa
and $V_{\mathrm{imp}}=42$ m/s (Fig.~\ref{figs5}).}
\label{figs6}
\end{figure}

Curves of propagation speed versus time for all five
cases analyzed are shown in Fig.~\ref{figs6}. For reference, the
longitudinal, $c_{l}$, shear, $c_{s}$, and $\sqrt{2}c_{s}$ ($=1775$ m/s)
wave speeds are marked. Also shown in the 
figure are data points corresponding to measured propagation speeds for
the experiment with $\Sigma_0=9.4$ MPa and an impact
velocity of 42 m/s. The first step in computing the propagation speeds 
is to record the location of the point furthest from the impact edge
where $\Delta \dot{u}_{\rm slip}>0$.
This gives the position of the leading sliding tip at various times and the
propagation speed, $V_{\rm tip}$, is then calculated from this data by
a progressive least squares fit using five points. At least some of
the oscillations in propagation speed in Fig.~\ref{figs6} may be due
to this numerical differentiation procedure.

Sliding generally
initiates somewhat away from the impact edge and then occurs over
several small sliding regions that link up. Thus, the
process of sliding initiation is not one of propagation of a sliding
region but involves jumps in the location of the leading sliding
edge. As a consequence, the `speeds' calculated during the early
stages of sliding are 
unphysically high. Subsequently, the propagation speeds decrease and
for Cases II and V, the mean $V_{\rm tip}$ is about $c_l$ whereas for
Case IV,  after the initial stages of sliding, $V_{\rm tip}$ reduces
to $\approx 2000$ m/s, in good agreement with the experimental value
of 1950 m/s. There are multiple pulses in Case IV and 
the tip of the second pulse travels at about 2700 m/s, as compared with 
$\approx 2600$ m/s in the experiments. 
It is worth emphasizing that both the experiments and the calculations
indicate that the second pulse travels at a supersonic speed. However,
for the calculation of Case IV, $\Sigma_0=0.9$ MPa and 
$V_{\rm imp}=10$ m/s as compared with 9.4 MPa and 42 m/s in the
experiments. Hence, there is consistency of propagation speeds for 
this mode of interface sliding although the loading conditions to achieve
this are quite different in the calculations and the
experiments. One possibility for this discrepency is that the effect of
the impedence mismatch between the steel plate bonded to the specimen
and Homalite is not accounted for in the
computations.

In the following, distributions are shown both for the evolution of
$\Delta \dot{u}_s$, 
which is the particle velocity jump, and for 
$\Delta \dot{u}_{\mathrm{slip}}$, which is the sliding rate obtained
from the frictional constitutive 
relation. From eq.~(\ref{eqtime1}), the difference between these is the
`elastic' sliding rate $\dot{T}_s/C_s$. There are cases where 
$\Delta \dot{u}_s \approx \Delta \dot{u}_{\mathrm{slip}}$ with the
elastic sliding rate 
being negligible. However, in other cases the difference between $\Delta 
\dot{u}_s$ and $\Delta \dot{u}_{\mathrm{slip}}$ is substantial and the
elasticity of the interface then plays a significant role.

The principal stress contours that evolve are a consequence of both
the imposed loading and sliding along the interface. To provide a
perspective on the contribution of interfacial sliding, the 
contours of $\sigma_1-\sigma_2$ that emerge from impact loading with
no sliding are considered first.

\subsection{Symmetric loading with respect to the interface}

\begin{figure}[h]
\begin{center}
\includegraphics[width=4.5in]{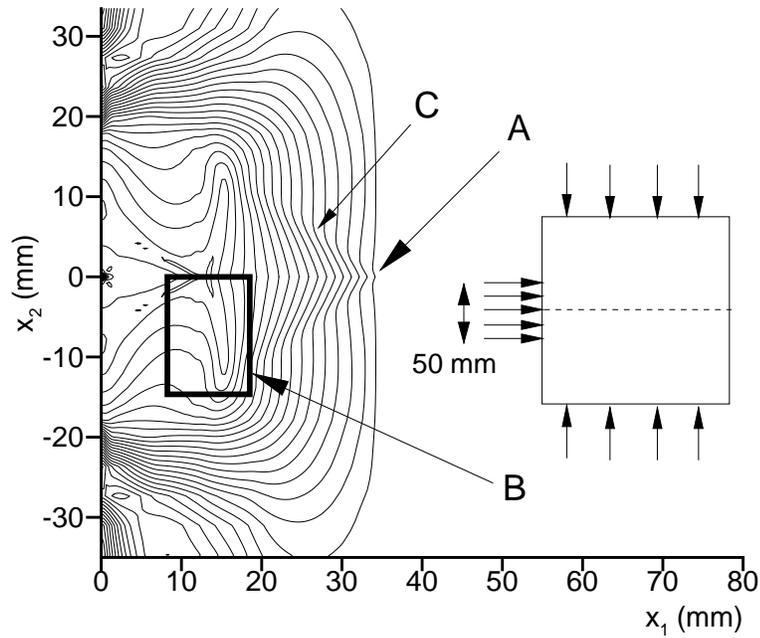} \\
(a) \\
\includegraphics[width=3.5in]{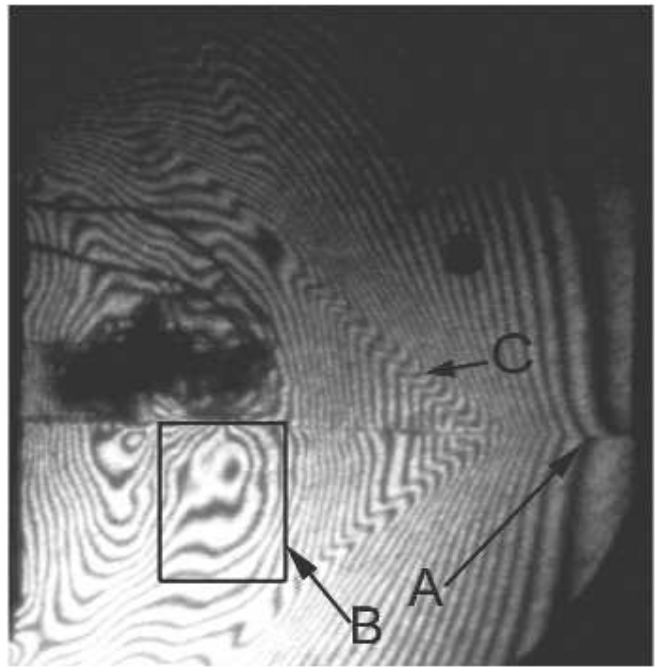} \\
(b) \\
\end{center}
\caption{Isochromatic fringe pattern due to symmetric
loading with respect to the interface. (a) Numerical contours at 
$t=16$ $\mu$s for $\Sigma_0=10$ MPa, $V_{\mathrm{imp}}=5$ m/s, with
the configuration 
analyzed shown in the inset, (b) Experimental isochromatic fringe
patterns for $\Sigma_0=0$ MPa, $V_{\mathrm{imp}}=58$ m/s. In (b)
the field of view is 130 mm in diameter.}
\label{fig7}
\end{figure}

In order to illustrate the isochromatic fringe patterns that result
without sliding, the symmetric loading configuration shown in the
inset of Fig.~\ref{fig7}a is analyzed.  Figure \ref{fig7}a shows the
simulated isochromatic fringe pattern at $t=16$ $\mu$s for a
calculation carried out with $\Sigma_0=10$ MPa and
$V_{\mathrm{imp}}=5$ m/s. At this time, the loading wave front has
propagated 34 mm into the specimen.  The isochromatic fringe pattern
for a symmetric loading experiment with $\Sigma_0=0$ MPa and
$V_{\mathrm{imp}}=58$ m/s is shown in Fig.~\ref{fig7}b.  In both the
experiments and simulations the continuity of fringes across the
interface behind the loading wave front indicates that no sliding
occurs along the interface under symmetric loading.

In both the simulations and the experiments, at the loading wave front
there is a cusp in the stress contours at the interface (marked by the
arrow A), indicating that the propagation speed is faster along the
interface than in the bulk. The cusp in the fringes at the interface
in Fig.~\ref{fig7}b is also seen when frictional sliding occurs
(Figs.~\ref{figs4} and \ref{figs5}).  Another characteristic feature is
marked by box B in Fig.~\ref{fig7}. The curvature of the fringe lines
associated with this feature is greater in the computations in
Fig.~\ref{fig7}a, but this curvature decreases with time. A similar
feature, with a curvature closer to what is seen in the computations, is
in Fig.~\ref{figs5}b.  Our simulations indicate that the cusp in
the fringe line and the feature in box B are a consequence of the
impact loading conditions and thus not directly associated with
frictional sliding.

A relatively broad head wave that emanates from the interface
following the loading wave front is shown by the arrow C.  The width
of the head wave in the simulations was found to depend on $C_n$ and 
$C_s$ in eq.~(\ref{eqtime1}) and thus can be used to set the values of
the interface elastic stiffnesses.

\subsection{Case I: Crack-like mode}

\begin{figure}[tbp]
\begin{center}
\includegraphics[width=4.5in]{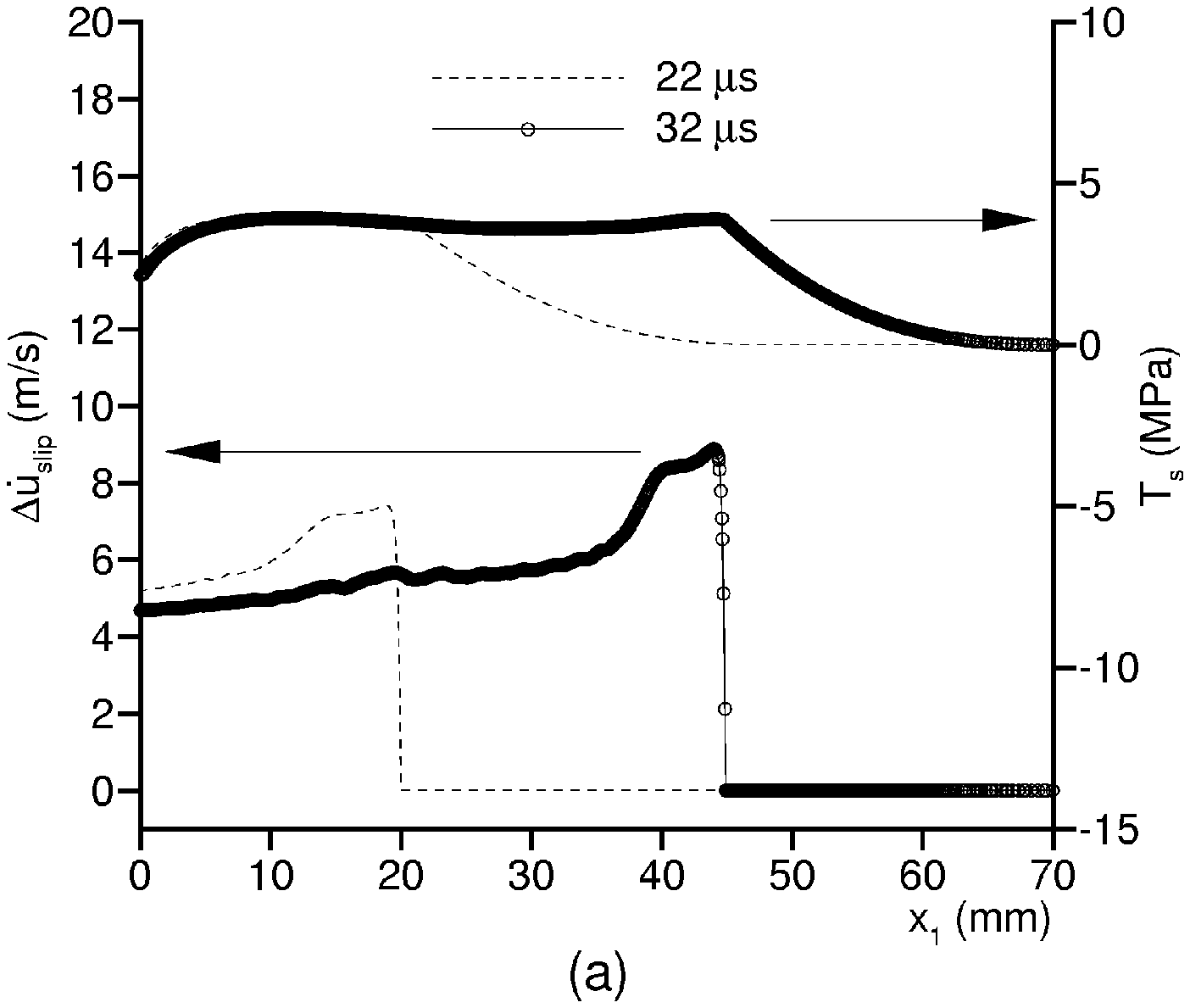} 
\includegraphics[width=4.5in]{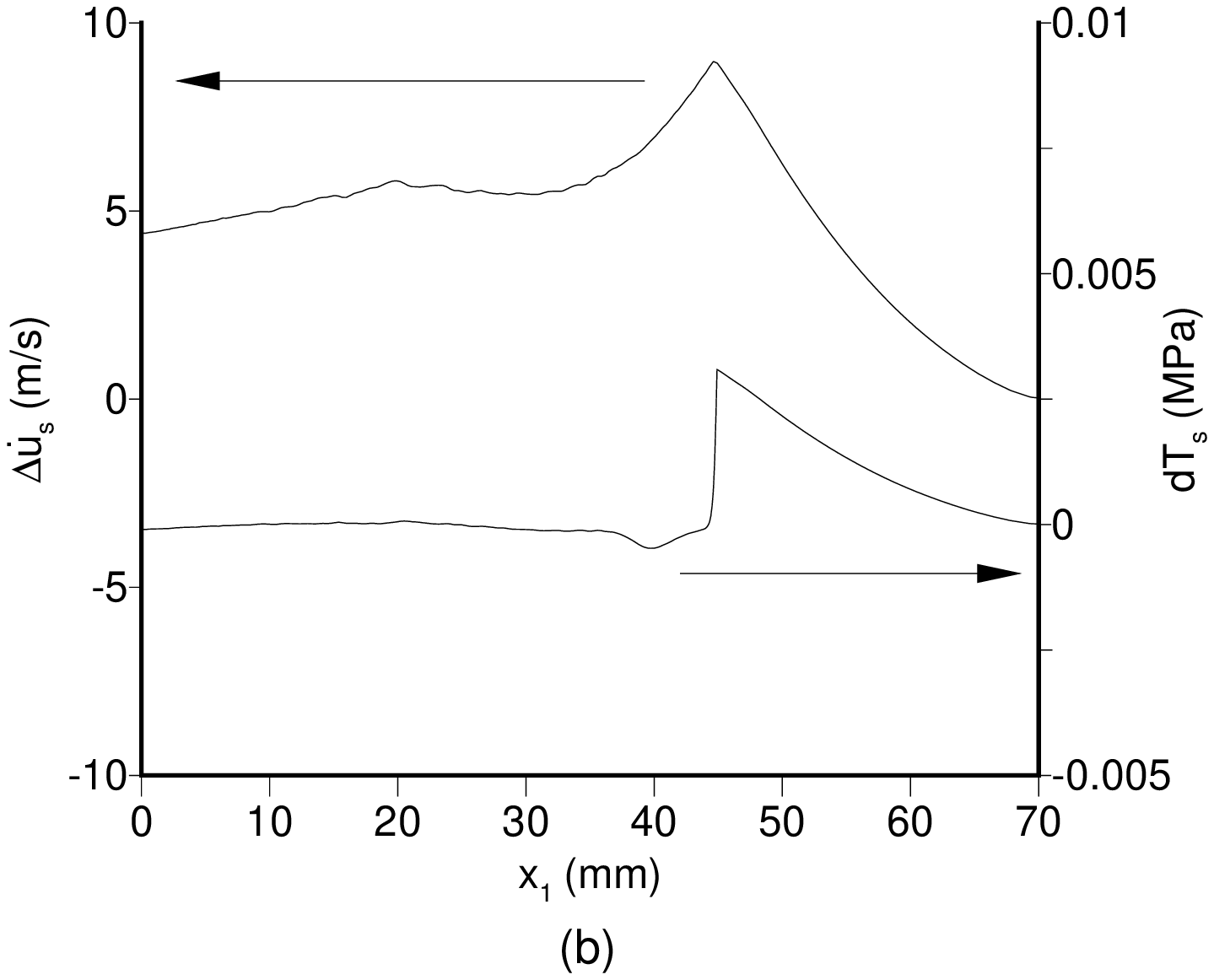}
\end{center}
\caption{Distributions along a portion of the interface for
Case I -- $\Sigma_0=6$ MPa, $V_{\mathrm{imp}}=2$ m/s. (a) The frictional
sliding rate, $\Delta \dot{u}_{\mathrm{slip}}$, and the shear traction, $T_s$
at $t=22$ $\mu$s and at $t=32$ $\mu$s. The symbols mark values at
integration points along the interface to illustrate the resolution of
the discretization. (b) The shear traction increment, 
$dT_s=\dot{T}_s dt$, and the velocity jump across the interface, 
$\Delta \dot{u}_{s}$, at $t=32$ $\mu$s.}
\label{fig8}
\end{figure}

Figure~\ref{fig8}a shows the variation of $\Delta
\dot{u}_{\mathrm{slip}}$, and the shear traction, $T_s$, along the 
interface at $t= 22$ $\mu$s (dashed lines) and at $t= 32$ $\mu$s (solid
lines) for Case I ($\Sigma_0=6$ MPa, $V_{\mathrm{imp}}=2$ m/s). Frictional
sliding initiates at the impact edge ($x_1=0$) $16$ $\mu$s after impact. The
distributions of $\Delta \dot{u}_{\mathrm{slip}}$ and $T_s$ along the
interface are similar at $22$ $\mu$s and $32$ $\mu$s, indicating at
least quasi-steady behavior. 

Ahead of the frictional sliding tip, $\Delta \dot{u}_{\mathrm{slip}}=0$.
At the tip, $\Delta \dot{u}_{\mathrm{slip}}$ jumps to its peak value of
about 7 m/s within two mesh spacings (eight interface integration
points) and then gradually reduces to $\approx 5$ 
m/s. As seen in Fig.~\ref{fig8}a, the peak value of 
$\Delta \dot{u}_{\mathrm{slip}}$ gradually increases with 
time. This behavior is crack-like in that frictional sliding persists at a
point on the interface after the sliding front has past that point.

The shear traction steadily increases from the loading wave front, which is
at $x_1=71$ mm at $t=32$ $\mu$s, and reaches a maximum value of about 3.9
MPa at the frictional sliding tip after which it decreases to a steady value
of 3.6 MPa. The normal traction along the interface gradually increases from
its initial value of 6 MPa to 6.39 MPa at the frictional sliding tip and
subsequently decreases to 6.14 MPa along the sliding region. The apparent
coefficient of friction, $\mu_{\mathrm{app}}=T_s/T_n$, is $\approx 0.609$ at
the frictional sliding tip and reduces to 0.585 along the sliding region
with small variations. Thus, in this case $\mu_{\mathrm{app}}$ remains very
close to $\mu_s$. The value of $T_n$ itself is nearly constant along
the interface, differing by less than $0.6$ MPa from $\Sigma_0=6$ MPa,
except near the impact edge ($0 \le x_1 \le 10$ mm) where the
magnitude of $T_n$ decreases.

The actual velocity jump across the interface is $\Delta \dot{u}_{s}$, which
consists of an elastic part in addition to $\Delta \dot{u}_{\mathrm{slip}}$.
Figure \ref{fig8}b shows the distribution of $\Delta \dot{u}_{s}$ as well as
the traction increment $dT_s=\dot{T}_s dt$ distribution. Because of the
elasticity in the interface constitutive relation, 
eq.~(\ref{eqtime1}), $\Delta \dot{u}_{s}$ increases gradually to it
peak value, rather than 
exhibiting a sharp jump. However, behind the frictional sliding tip, $\Delta 
\dot{u}_{s}$ and $\Delta \dot{u}_{\mathrm{slip}}$ are nearly equal.
Consistent with this, since the elastic contribution is 
$\dot{T}_s/C_s$, $\dot{T}_s \approx 0$ behind the frictional sliding
tip. 

\begin{figure}[tbp]
\begin{center}
\includegraphics[width=5in]{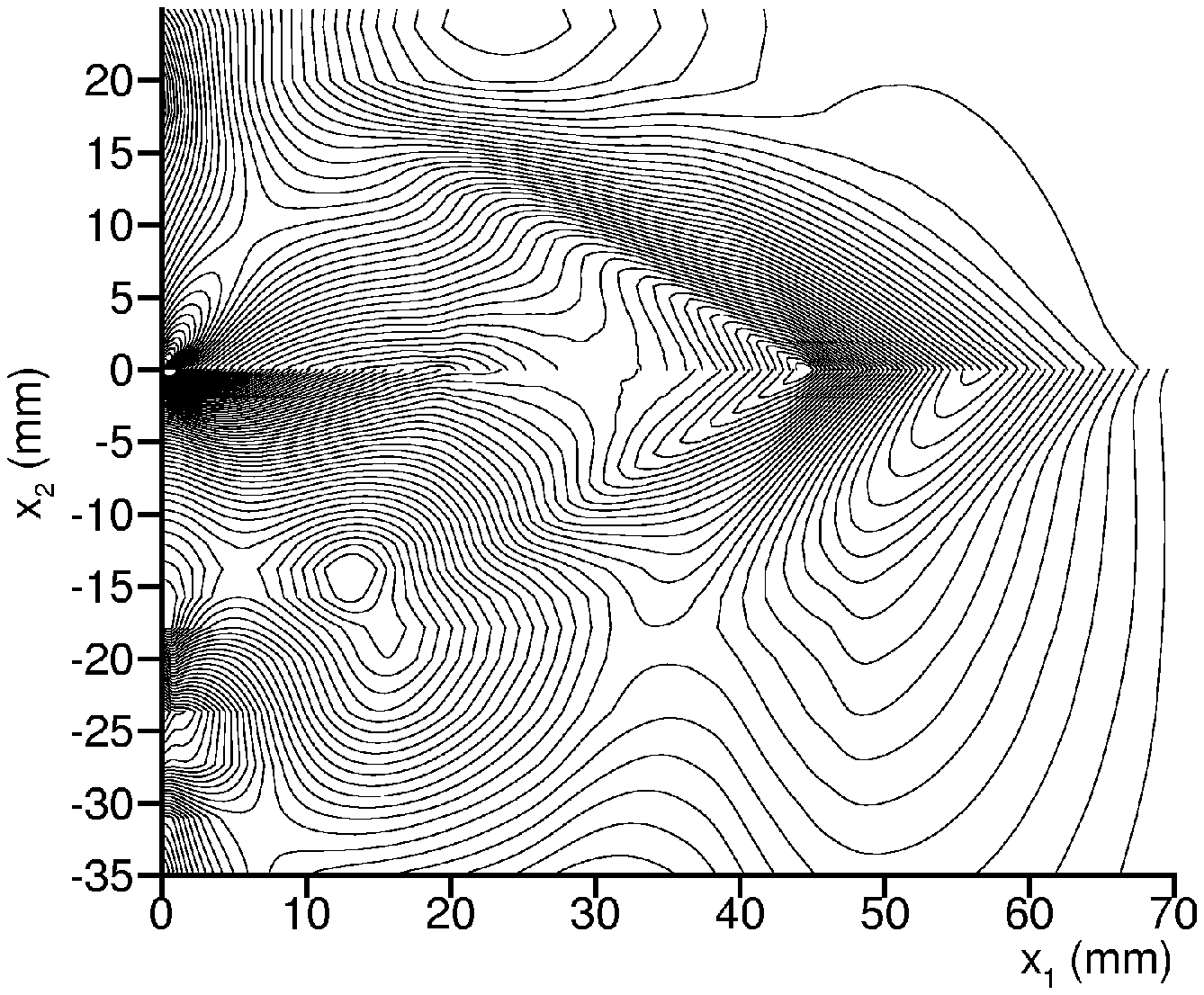}
\end{center}
\caption{Contours of $\sigma_1-\sigma_2$ (isochromatic
fringe patterns) at $t=32$ $\mu$s for Case I.}
\label{figs9}
\end{figure}

Isochromatic fringe patterns (contours of $\sigma_1-\sigma_2$) at
$t=32$ $\mu$s are shown in Fig.~\ref{figs9}. Behind the loading wave
front (which is at  
$71$ mm) a concentration of fringes at 44 mm indicates the location of the
frictional sliding tip. A shear Mach cone, formed by a sharp change in
fringe density, emanates from the sliding tip at an angle of 34$^\circ$.
Using $\Delta a/\Delta t =\sin\theta/c_s$, where $\theta$ is the Mach angle,
gives a tip speed of 2244 m/s which is close to, but less than, the
directly calculated tip speed in Fig.~\ref{figs6}. Another Mach cone,
formed by a more 
gradual change in fringe density at an angle of 67$^\circ$ and, presuming
that $c_l$ is the relevant speed of sound for this Mach cone, a propagation
speed of 2360 m/s is obtained from ${\Delta a/\Delta t}=\sin\theta/c_l$.
Behind the frictional sliding tip there is no significant stress
concentration, indicative of sliding occurring along the interface.

\subsection{Case II: Pulse-like mode}

\begin{figure}[tbp]
\begin{center}
\includegraphics[width=4.5in]{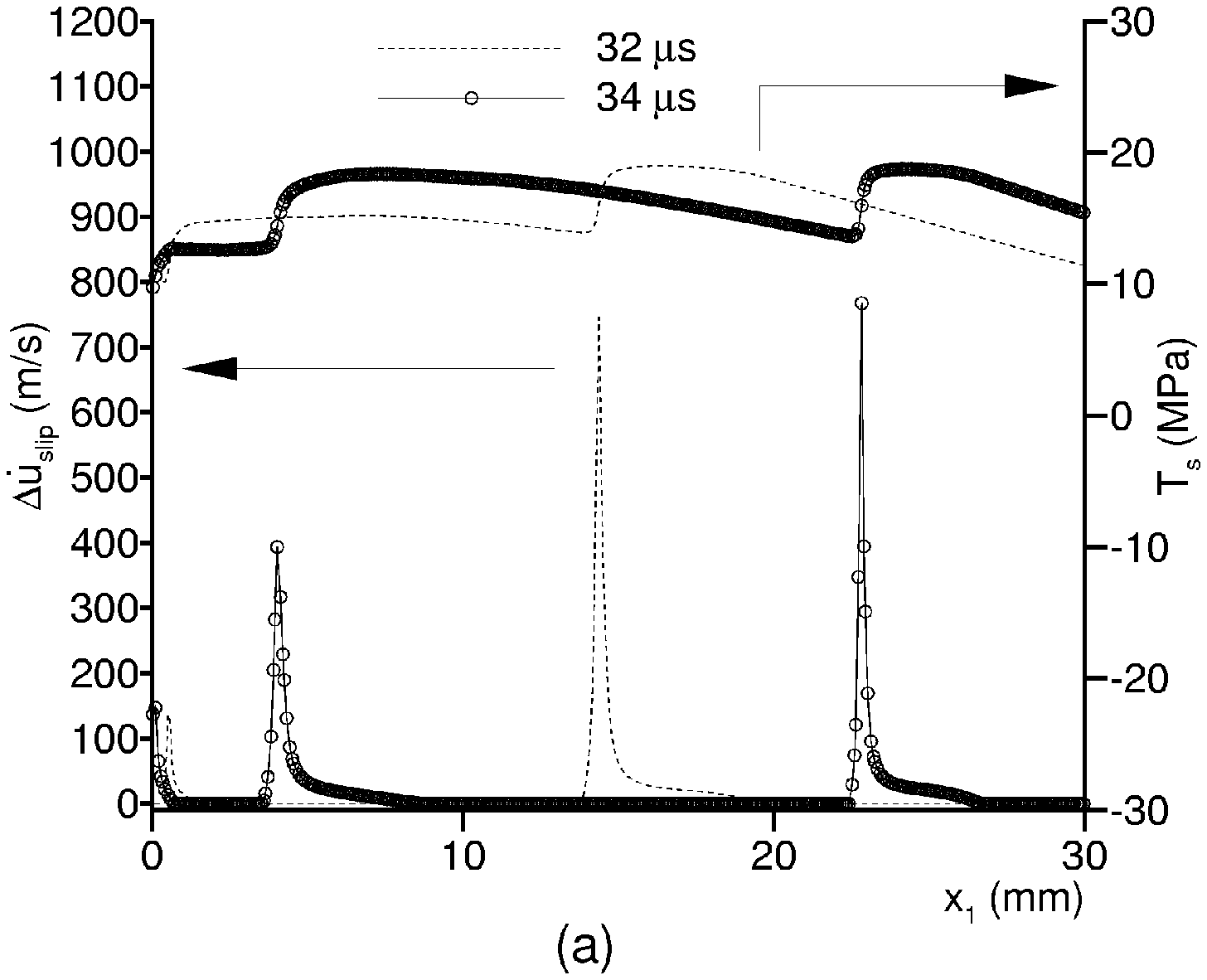} 
\includegraphics[width=4.5in]{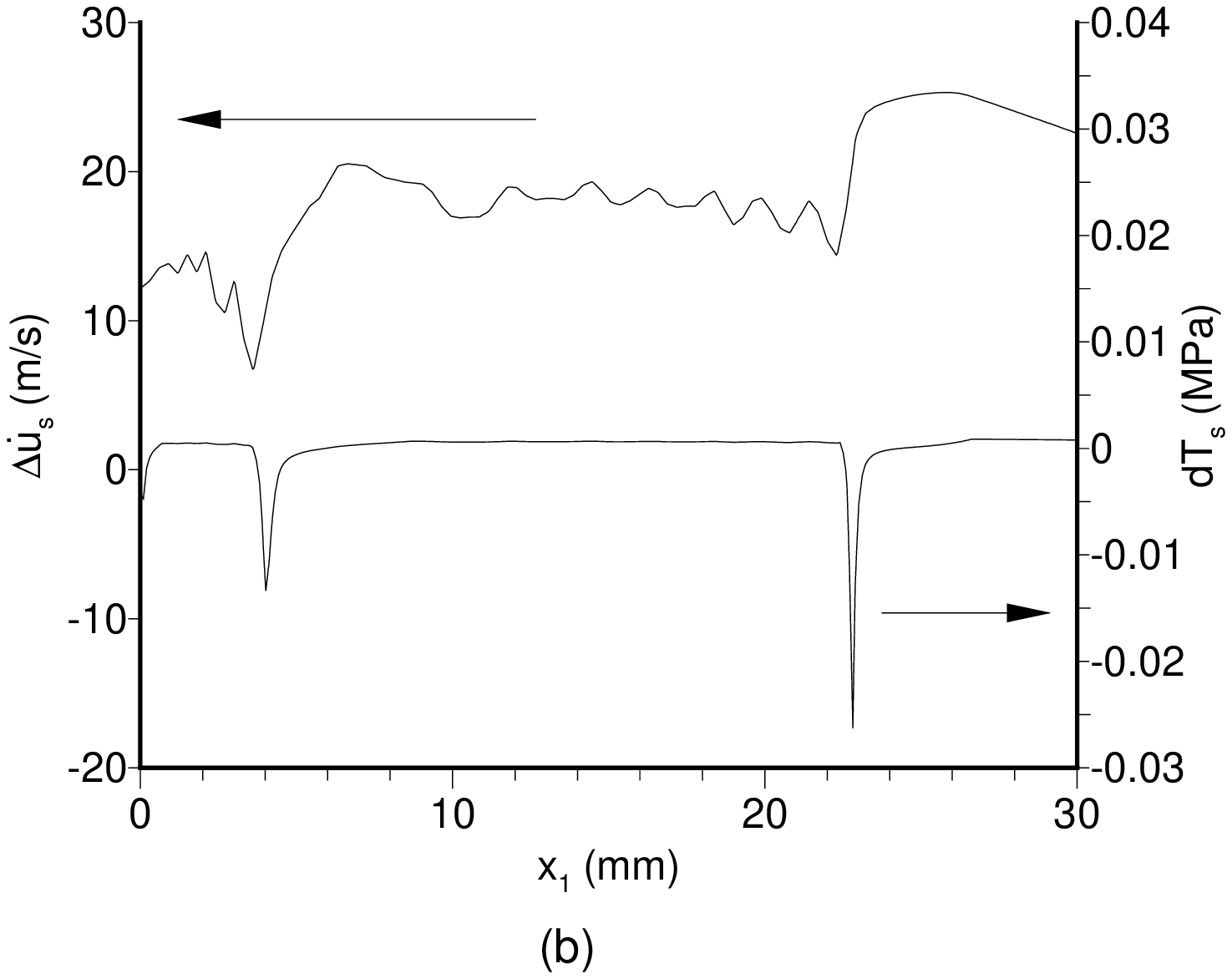}
\end{center}
\caption{Distributions along a portion of the interface for
Case II -- $\Sigma_0=30$ MPa, $V_{\mathrm{imp}}=2$ m/s. (a) The frictional
sliding rate, $\Delta \dot{u}_{\mathrm{slip}}$, and the shear traction, $T_s$
at $t=32$ $\mu$s and at $t=34$ $\mu$s. The symbols mark values at
integration points along the interface to illustrate the resolution of
the discretization. (b) The shear traction increment, 
$dT_s=\dot{T}_s dt$, and the velocity jump across the interface,
$\Delta \dot{u}_{s}$, at $t=34$ $\mu$s.}
\label{figs10}
\end{figure}

In Fig.~\ref{figs10}a, the variation of the frictional sliding rate, $\Delta 
\dot{u}_{\mathrm{slip}}$, and the shear traction, $T_s$, along the interface
at $t=32$ $\mu$s (dashed
lines) and at $t= 34$ $\mu$s (solid lines with circles)
are shown for Case II ($\Sigma_0=30$ MPa, $V_{\mathrm{imp}}=2$ m/s).
Frictional sliding initiates at $t=28$ $\mu$s, which is later than for Case
I because the larger compressive stress requires a larger shear stress for
initiation. In Case II, frictional sliding occurs in a pulse-like mode which
involves a relatively narrow zone (the pulse) where 
$\Delta \dot{u}_{\mathrm{slip}}>0$, with $\Delta
\dot{u}_{\mathrm{slip}}=0$ ahead of and behind the 
pulse. The circles in Fig.~\ref{figs10}a show the integration point locations
(since $\Delta \dot{u}_{\mathrm{slip}}$ is a constitutive quantity it is
stored at interface element integration points) to indicate the extent to
which the pulse is resolved; the sharp pulse consists of 9-10 points.

A pulse initiates at the impact edge and as it travels across the interface
the peak value of $\Delta \dot{u}_{\mathrm{slip}}$
increases. The pulse
profile consists of $\Delta \dot{u}_{\mathrm{slip}}$ gradually
increasing from zero to 40 m/s in 4.5 mm after which it 
sharply rises to 700 m/s over a distance of 0.5 mm and then drops abruptly
to zero over a distance of 0.4 mm.
Additional pulses initiate and evolve in a similar manner;
there are 
two pulses at $t=32$ $\mu$s and three at $t=34$ $\mu$s. 

The shear traction distribution along the interface is also 
shown in Fig.~\ref{figs10}a (note that the range for $T_s$ is larger
than in Fig.~\ref{fig8}a). At $t=32$ $\mu$s, the shear traction along
the interface gradually 
increases from zero at the loading wave front (which is at $71$ mm) to 18
MPa at the tip of the leading pulse. The value of $T_s$ remains
approximately constant over the pulse, abruptly decreases to 14 MPa
at the trailing edge of the pulse and then very gradually increases until
the second slip-pulse is reached. The variation of the apparent coefficient
of friction, $\mu_{\mathrm{app}}=T_s/T_n$, is quite different from that for
the crack-like mode. Here, at $t=32$ $\mu$s, $\mu_{\mathrm{app}}$ is 0.60 at
the front of the frictional sliding tip, very slightly increases to 0.61
over the pulse and then, with the drop in $T_s$, falls to 0.44. In this
case, neither $\mu_s$ nor $\mu_d$ serve as bounds on the apparent
coefficient of friction. Behind the leading pulse the apparent coefficient
of friction steadily increases to 0.60 until the trailing pulse is
reached. The magnitude of $T_n$ is nearly constant at $\Sigma_0=30$
MPa over most of the interface but decreases near the impact edge ($0
\le x_1 \le 10$ mm).

In contrast to the crack-like mode, for the pulse-like mode there is a
qualitative difference between the distributions of 
$\Delta \dot{u}_{\mathrm{slip}}$ and $\Delta \dot{u}_{s}$ along the
interface. As seen in Fig.~\ref 
{figs10}b, which shows distributions of $\Delta \dot{u}_{s}$ 
and $d{T}_s=\dot{T}_s dt$ at $t=34$ $\mu$s, no pulse occurs in the
distribution of $\Delta  
\dot{u}_{s}$. However, there is an abrupt change in $\Delta \dot{u}_{s}$ at
the location of each pulse. On the other hand, there are pulses in
the traction increment $d {T}_s$ and hence in the 
elastic contribution to $\Delta \dot{u}_{s}=\dot{T}_s/C_s$. The abrupt
change in $\Delta \dot{u}_{s}$ 
corresponds to a pulse in $d\Delta \dot{u}_{s}/dx$, i.e. to a `weak' pulse
as opposed to the `strong' pulse in $\Delta \dot{u}_{\mathrm{slip}}$.

\begin{figure}[tbp]
\begin{center}
\includegraphics[width=5in]{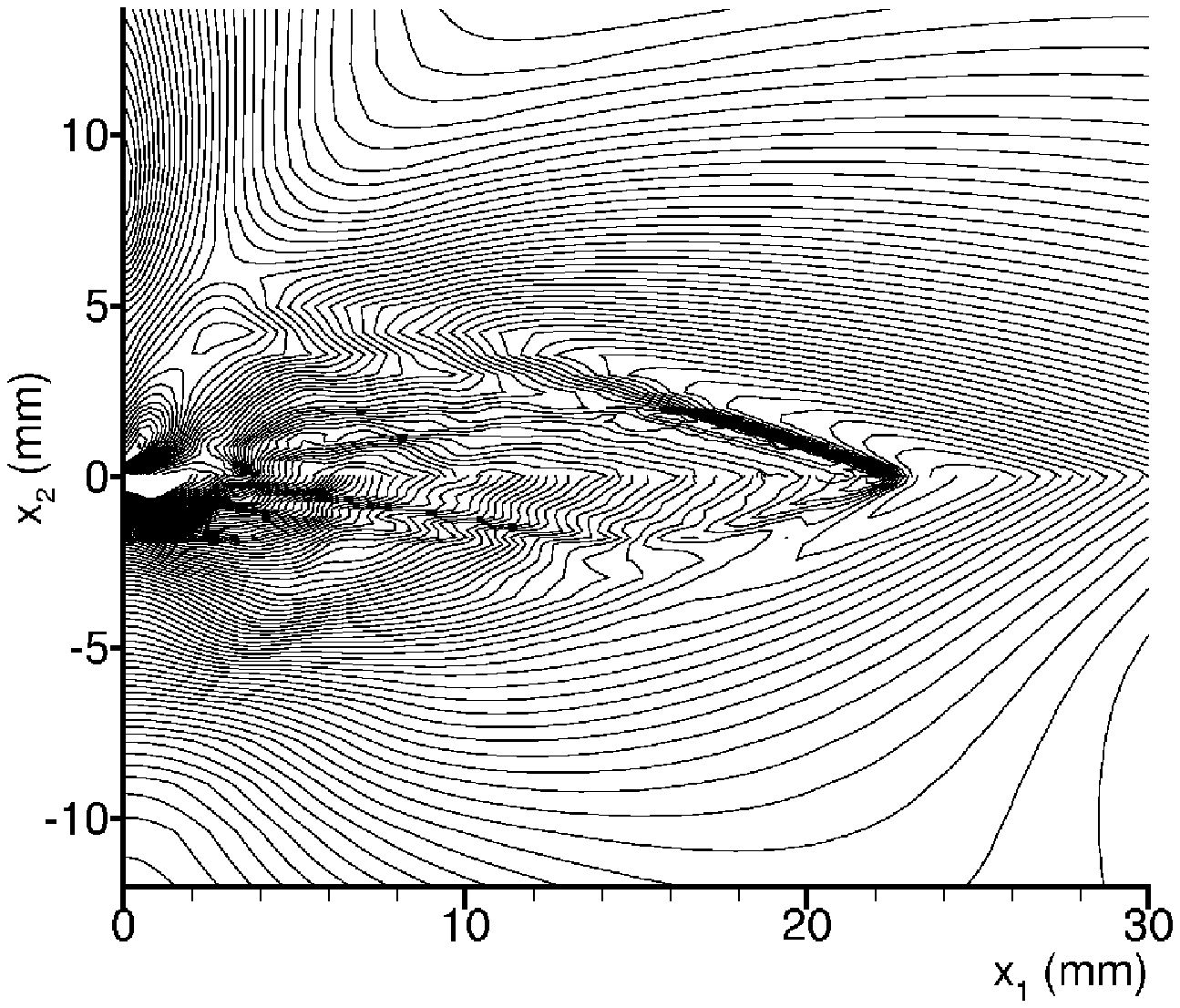}
\end{center}
\caption{Contours of $\sigma_1-\sigma_2$ (isochromatic
fringe patterns) at $t= 34$ $\mu$s for Case II.}
\label{figs11}
\end{figure}

Isochromatic fringe patterns (contours of $\sigma _{1}-\sigma _{2}$) at $t=32
$ $\mu $s are shown in Fig.~\ref{figs11} where only the stress contours
behind the loading wave front are visible in the region shown. There is a
concentration of fringes at approximately $x_{1}=22$ mm, which
corresponds to the location of the leading slip pulse. The change in the
fringe spacing and concentration along a shear Mach line of $26^{\circ }$
indicates a propagation speed of about 2800 m/s ($\sin \theta
/c_{s}$). The dense set of 
fringe near the impact site arises from a slip pulse initiating there.

\subsection{Case III: A train of pulses}

\begin{figure}[tbp]
\begin{center}
\includegraphics[width=4.5in]{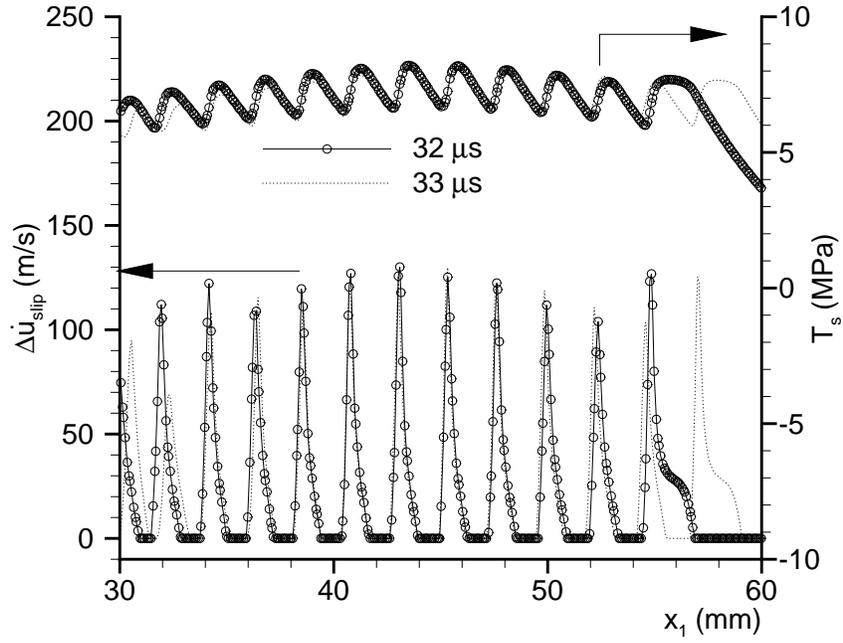} 
\includegraphics[width=4.5in]{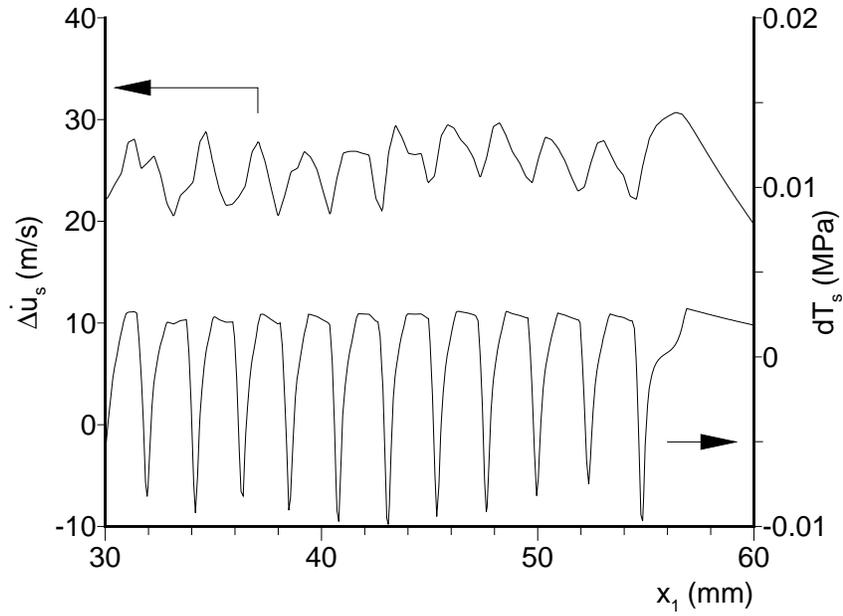}
\end{center}
\caption{Distributions along a portion of the interface
for Case III -- $\Sigma_0=10$ MPa, $V_{\mathrm{imp}}=20$ m/s. (a) The
frictional sliding rate, $\Delta \dot{u}_{\mathrm{slip}}$, and the shear
traction, $T_s$ at $t=32$ $\mu$s and at $t=33$ $\mu$s. (b) The shear
traction increment, $dT_s=\dot{T}_s dt$, and the velocity jump across the
interface, $\Delta \dot{u}_{s}$, at $t=32$ $\mu$s.}
\label{figs12}
\end{figure}

Figure \ref{figs12}a shows distributions of frictional sliding rate, $\Delta 
\dot{u}_{\mathrm{slip}}$, and shear traction $T_s$ for a train of 
pulses ($\Sigma_0=10$ MPa, $V_{\mathrm{imp}}=20$ m/s). The
distributions are shown at two times, at $t= 32$ $\mu$s and at $t=
33$ $\mu$s. As in Fig.~\ref{figs10}a, the circles indicate the positions
of the integration points at which $\Delta \dot{u}_{\mathrm{slip}}$
and $T_s$ are evaluated. A key difference 
between the pulses in Case III and those in Case II is that in Case III the
peak value of $\Delta \dot{u}_{\mathrm{slip}}$, which is $\approx 120$ m/s,
does not increase with propagation distance or time. The variations in the
peak value of $\Delta \dot{u}_{\mathrm{slip}}$ seen in Fig.~\ref{figs12}a is
mainly due to some pulses attaining their peak value between the plotted
integration points. Typically, a pulse extends over four elements
(sixteen integration points). The slip pulses at 33 $\mu$s (dashed
lines) lie on top of the pulses at 32 $\mu$s with a one pulse offset,
illustrating the self-similar propagation of these pulses. 

The shear traction along the interface has a mean value of about 7 MPa with
oscillations of 1.5 MPa. The apparent coefficient of friction, 
$\mu_{\mathrm{app}}=T_s/T_n$ oscillates between 0.59 and 0.48 for the
slip-pulses behind 
the leading pulse, except for the leading pulse where the maximum is 0.63.
The value of $T_n$ is fairly uniform over the interval shown, having a
maximum of about 14 MPa at $x_1 \approx 45$ mm and decreasing smoothly
to around $11.5$ MPa at $x_1=30$ mm and $x_1=60$ mm, so that the
variation in $\mu_{\mathrm{app}}$ arises principally from the oscillations
in $T_s$. Due to the direct effect in the friction relation, there is a lag
between the onset and cessation of frictional sliding and the maximum and
minimum values of $\mu_{\mathrm{app}}$. The maximum value of 0.59 for 
$\mu_{\mathrm{app}}$ is reached 0.43 mm behind the sliding tip
for the pulses following the leading pulse and the minimum value of 0.48
occurs 0.25 mm behind the location where $\Delta
\dot{u}_{\mathrm{slip}}$ peaks (for all pulses, including the leading
pulse). 

As also seen in Fig.~\ref{figs12}a, except for the leading pulse, the shape
of the pulses is the same, where the width of the frictional sliding
region ($\Delta \dot{u}_{\mathrm{slip}}>0$) is 1.33 mm and the width of
the region with $\Delta \dot{u}_{\mathrm{slip}}=0$ is 0.92 mm. The
spacing between the pulses is 
2.21 mm. The leading pulse has more structure to its front edge, and this
structure is preserved as the pulse propagates.

The velocity jump across the interface, $\Delta \dot{u}_{s}$, and the
traction increment $d T_s = \dot{T}_s dt$ are shown in Fig.~\ref{figs12}b.
The traction increment distribution reflects the pulse structure of $\Delta 
\dot{u}_{\mathrm{slip}}$, while the oscillations in $\Delta \dot{u}_{s}$ are
smoother.

\begin{figure}[tbp]
\begin{center}
\includegraphics[width=5in]{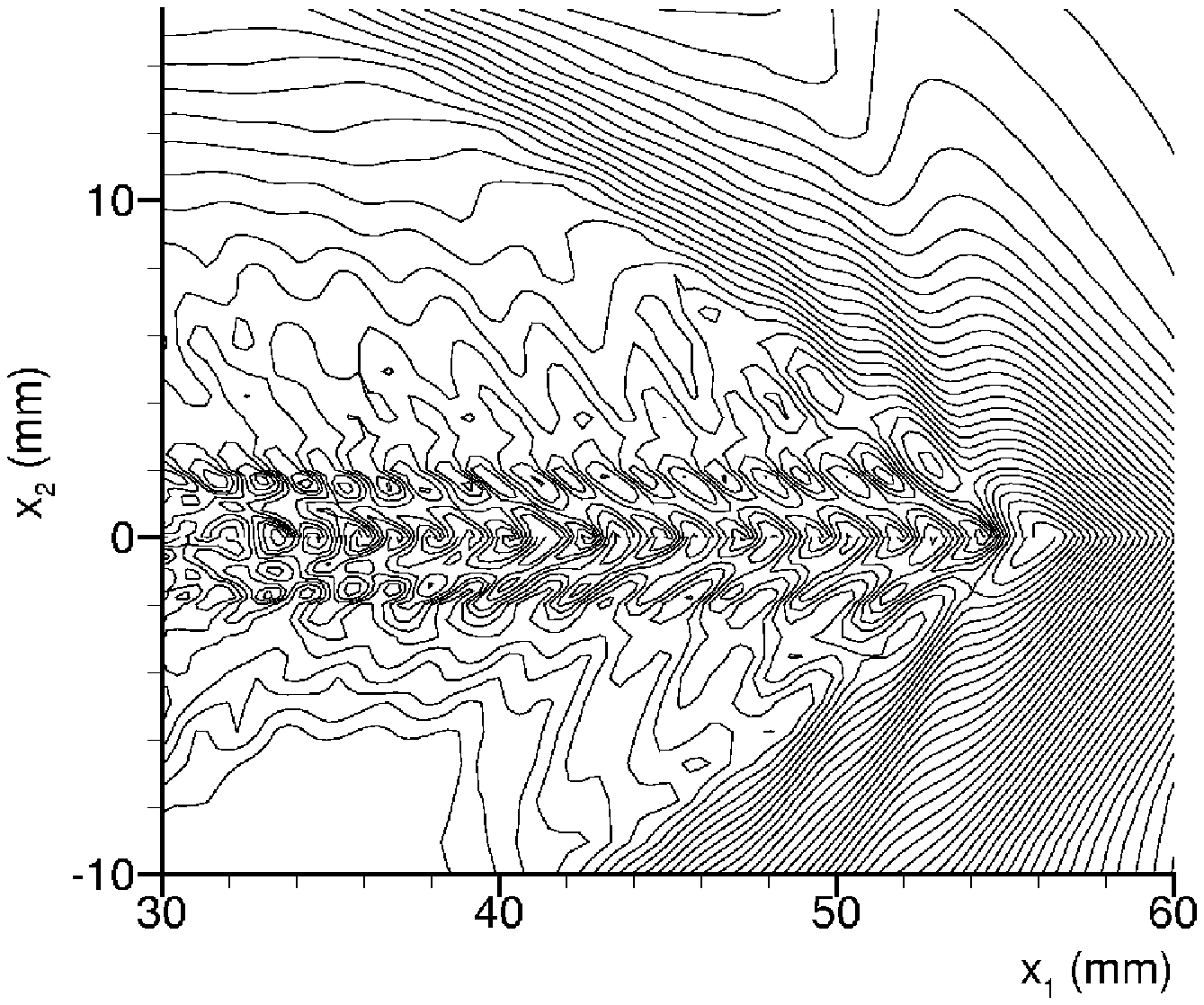}
\end{center}
\caption{Contours of $\sigma_1-\sigma_2$ (isochromatic
fringe patterns) at $t= 32$ $\mu$s for Case III.}
\label{figs13}
\end{figure}

Figure~\ref{figs13} shows the isochromatic fringe patterns 
(contours of $\sigma_1-\sigma_2$) at $t=32$ $\mu$s. At the front of
the first pulse at 55 
mm, a shear Mach wave occurs at $37^\circ$, corresponding to a propagation
speed of $\approx c_l$. Behind the slip front vortex like stress
contours are seen that correspond to the back and front of each pulse
together with shear Mach waves at each transition from slip to stick. Behind
the propagating front, there is a vortex-like structure. The change in the
contours at $x_2=2$ mm is due to the change in resolution of field
quantities associated with the transition from the uniform mesh to the
coarser, graduated mesh. This calculation was also carried out for a finer
mesh in which the uniform mesh region extends to $x_2=4$ mm, with the
element size in the uniform region fixed. The results are essentially
unchanged, but the finer mesh results in the details of the stress
distribution behind the propagating front being visible over the larger
region resolved.

\begin{figure}[tbp]
\begin{center}
\includegraphics[width=4.5in]{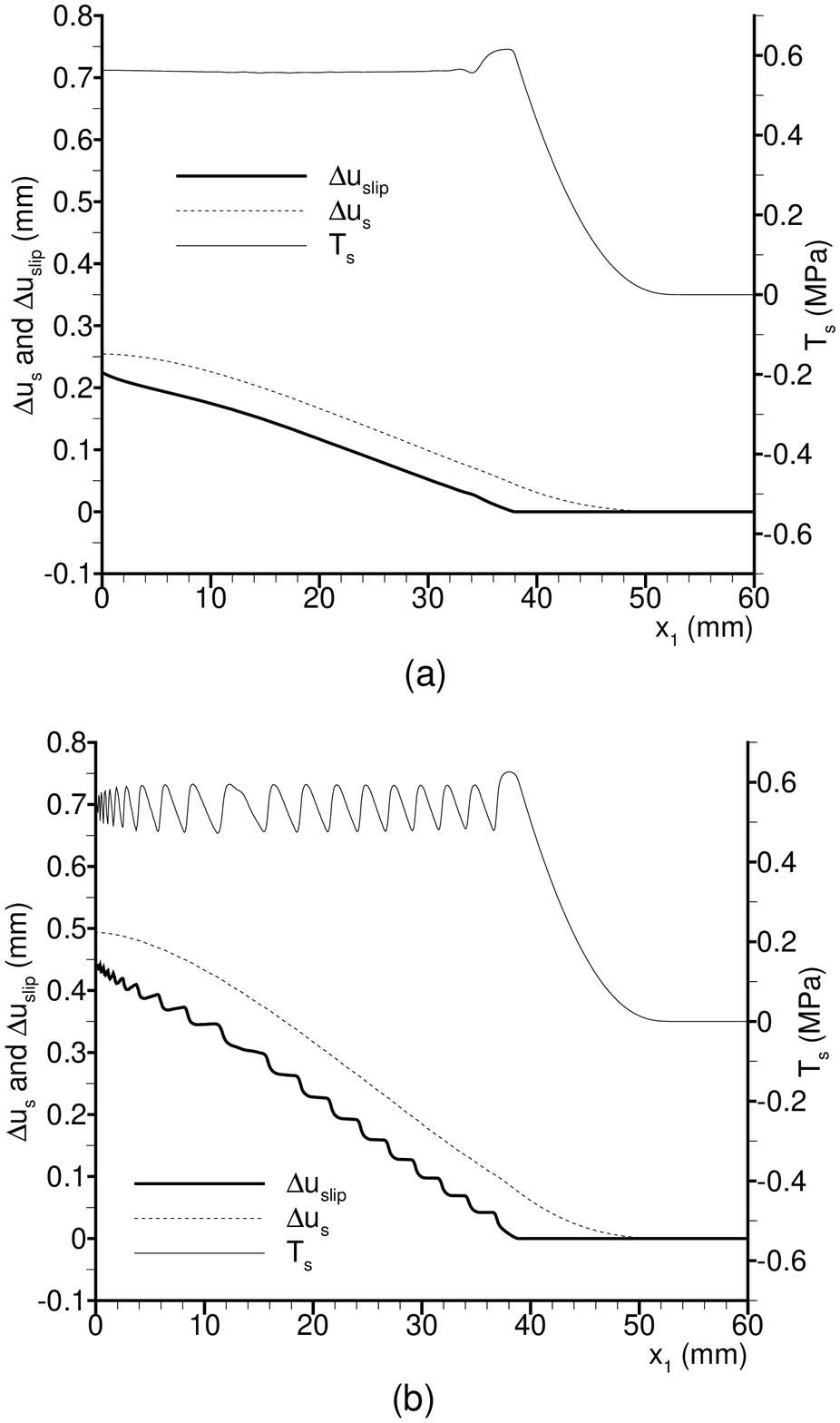}
\end{center}
\caption{Distributions of the accumulated frictional
sliding, $\Delta u_{\mathrm{slip}}=\int \Delta \dot{u}_{\mathrm{slip}} dt$,
the displacement jump across the interface, $\Delta u_s$ and the apparent
coefficient of friction, $\mu_{\mathrm{app}}=T_s/T_n$, across the interface.
(a) Case I -- $\Sigma_0=6$ MPa, $V_{\mathrm{imp}}=2$ m/s. 
(b) Case III -- $\Sigma_0=10$ MPa, $V_{\mathrm{imp}}=20$ m/s.}
\label{figs14}
\end{figure}

The accumulated frictional sliding, $\Delta u_{\mathrm{slip}}=\int_0^t
\Delta \dot{u}_{\mathrm{slip}}dt$, the displacement jump across the
interface, $\Delta u_s$ and the shear stress $T_s$ distributions for the
crack-like propagation mode (Case I) and for the train of pulses (Case III)
are compared in Fig.~\ref{figs14}. In both cases, the difference 
between $\Delta u_s$ and $\Delta u_{\mathrm{slip}}$ is due to the
assumed interface 
elasticity. In Fig.~\ref{figs14}a for Case I, both $\Delta u_s$ and $\Delta
u_{\mathrm{slip}}$ vary smoothly along the slipped surface. For Case III,
Fig.~\ref{figs14}b, although the displacement jump itself varies smoothly,
steps in $\Delta u_{\mathrm{slip}}$ are evident. The elasticity smooths out
the steps in $\Delta u_s$ along the interface, but these are reflected
in the 
oscillatory distribution of shear traction, $T_s$, along the interface.

\subsection{Case IV: Multiple pulses coalescing to form a crack}

\begin{figure}[tbp]
\begin{center}
\includegraphics[width=4.5in]{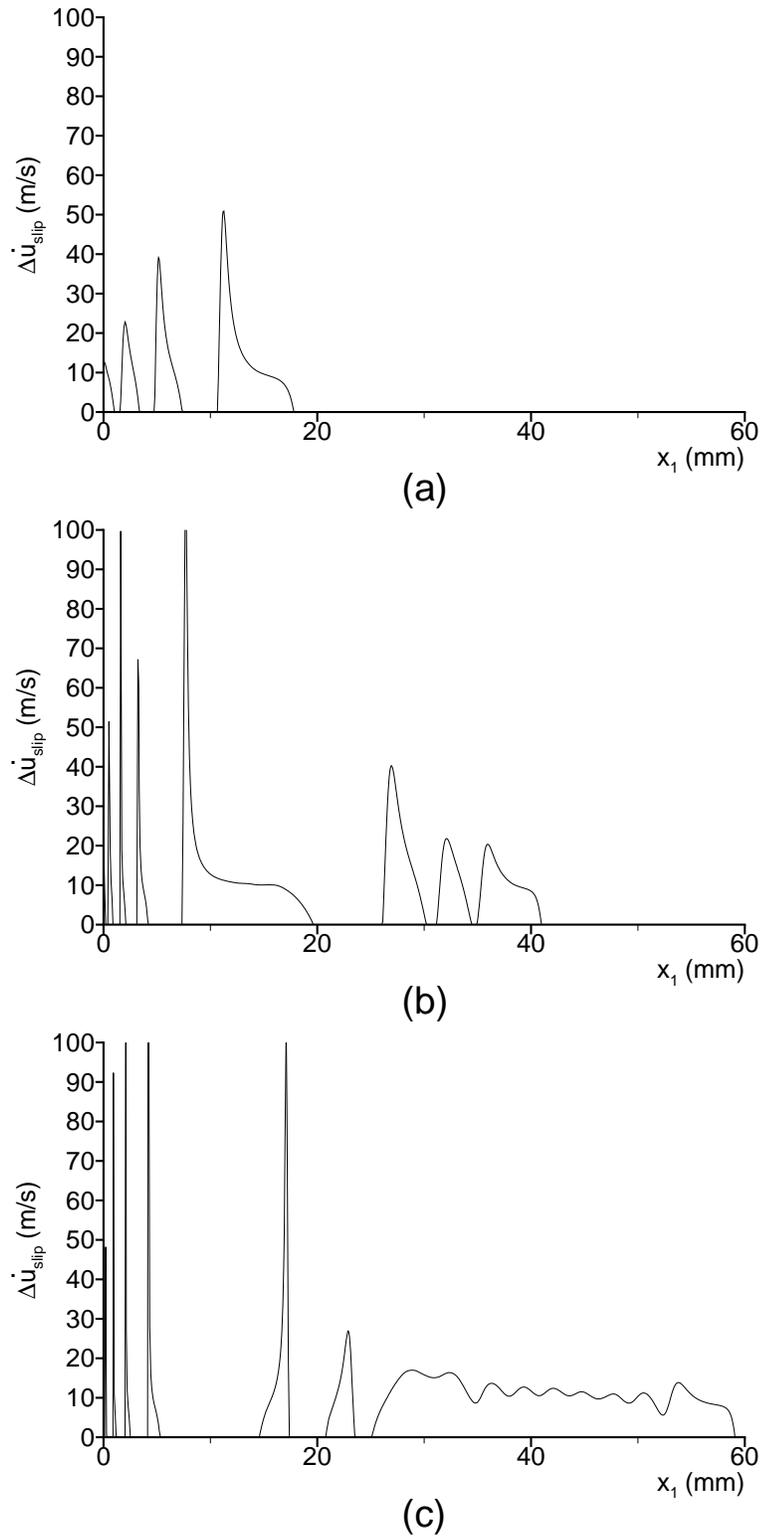}
\end{center}
\caption{Distribution of the frictional sliding rate, 
$\Delta \dot{u}_{\mathrm{slip}}$ for Case IV -- $\Sigma_0=0.9$ MPa, 
$V_{\mathrm{imp}}=10$ m/s. Also, for Case IV $C_n=0.03$ MPa/m and
$C_s=0.01$ MPa/m. 
(a) $t=22$ $\mu$s. (b) $t=32$ $\mu$s. (c) $t=48$ $\mu$s.}
\label{fig15}
\end{figure}

In Case IV, the values of the interface elastic constants
are $C_n=0.03$ MPa/m and $C_s=0.01$ MPa/m, one tenth their value
in all the other cases. Figure \ref{fig15} shows the slip velocity $\Delta 
\dot{u}_{\mathrm{slip}}$ along the interface at 22 $\mu$s, 32 $\mu$s,
and 41 $\mu$s. 
Sliding initiates at 14 $\mu$s and at 22 $\mu$s (Fig.~\ref{fig15}a) three
pulses are observed. The tip of the leading pulse at 18 mm and the distance
between the end of the first pulse and the front of the second pulse is 3.4
mm. As in Cases II and III, the leading pulse includes a gradual rising part
in the front before there is a steep rise in $\Delta \dot{u}_{\mathrm{slip}}$
whereas the second and third pulses do not exhibit such a gradual rise
in $\Delta \dot{u}_{\mathrm{slip}}$. At 32 $\mu$s in
Fig.~\ref{fig15}b, the tip 
of the leading pulse has moved to 41 mm with the distance between the first
and second pulse decreasing to 0.4 mm. During this time the average speed of
the tip of the leading pulse is $\approx c_l$ and the average speed of
the second pulse is supersonic, $\approx 2700$ m/s.
Thus, the second and third sliding pulses which are traveling
at supersonic speeds catch up with the lead pulse and coalesce to
form a single crack-like sliding region as shown in 
Fig.~\ref{fig15}c (at 41 $\mu$s) and then continue to travel at $c_l$.
This crack-like region expands with time as more pulses are generated at the
impact edge, travel at supersonic speeds and coalesce with the main
sliding region.

\begin{figure}[tbp]
\begin{center}
\includegraphics[width=4.5in]{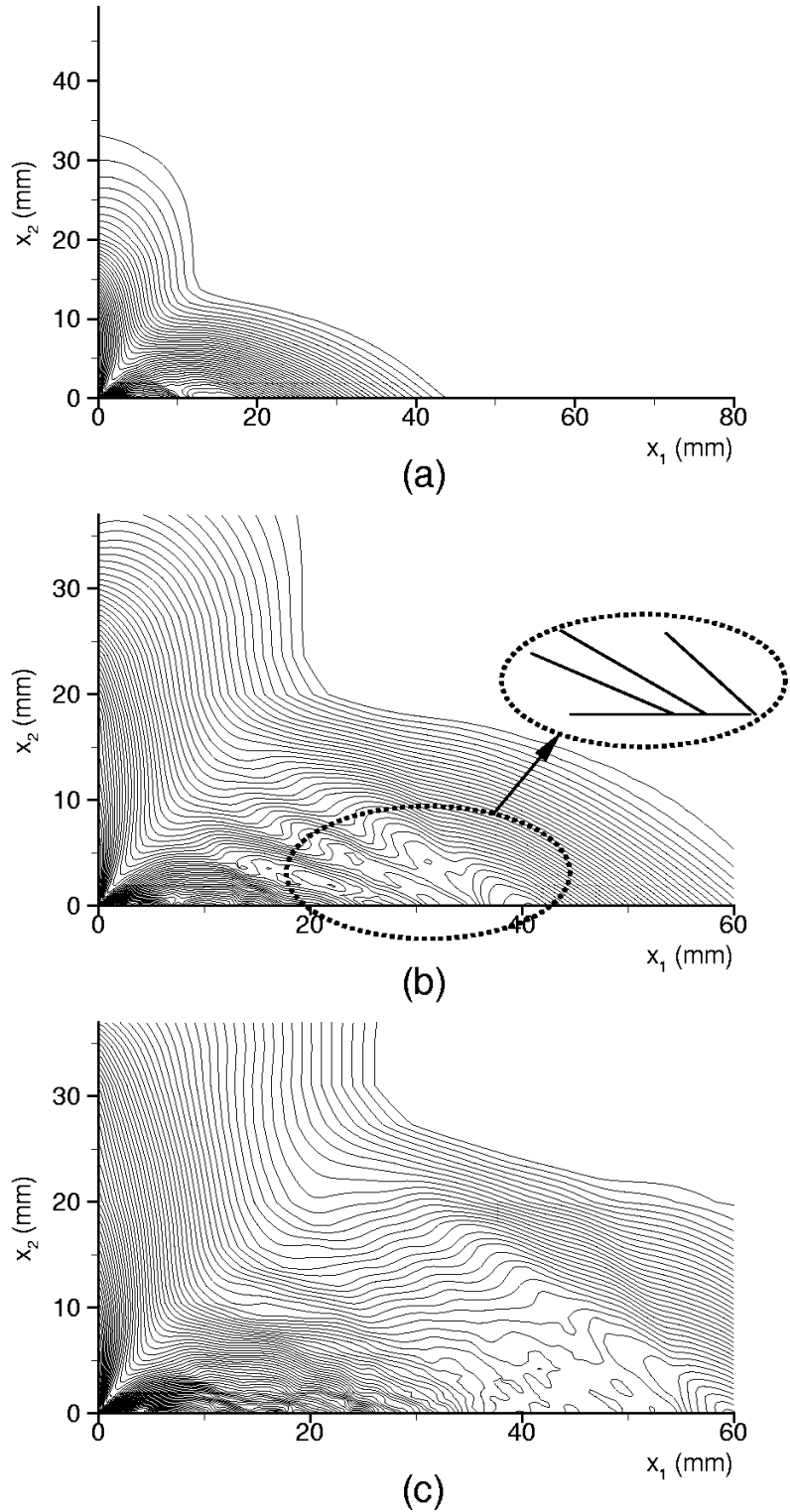}
\end{center}
\caption{Contours of $\sigma_1-\sigma_2$ (isochromatic
fringe patterns) for Case IV. The contour lines are drawn to focus on
details of the distribution for $x_2>0$. (a) $t=22$ $\mu$s. (b) $t=32$
$\mu$s. (c) $t=48$ $\mu$s. The inset 
in (b) indicates the Mach line orientations.}
\label{fig16}
\end{figure}

Isochromatic fringe patterns (contours of $\sigma_1-\sigma_2$) are shown in
Fig.~\ref{fig16}a, b, and c, at 22 $\mu$s, 32 $\mu$s, and 41 $\mu$s,
respectively. The contours are shown for $x_2>0$ because the Mach line are 
clearer in this plate. At $t=22 \mu$s the tip of the sliding
pulses is at 18 mm as shown by the change in the spacing of 
the stress contours
in Fig.~\ref{fig16}a.  In Fig.~\ref{fig16}b, the tip has moved to 41 mm.
There are three regions where an abrupt change in the density of the 
contours takes place forming three shear Mach lines at three 
angles to the interface as shown in the inset.  The intersections of these
lines with the interface are at the locations of the tips of
the three sliding pulses shown in  Fig.~\ref{fig16}b. The shear Mach angles
of 34$^\circ$, 29.3$^\circ$ and 23.4$^\circ$ correspond  to
propagation speeds of 2190 m/s, 2560 m/s and 
3160 m/s for the first, second  and third pulses, respectively. In Fig.~\ref
{fig16}c, at 41 $\mu$s,  the stress  contours show a concentration of
fringes only at 60 mm  corresponding to the sliding tip. At this time the
second and  third pulses have coalesced with the first pulse to form a 
crack-like sliding region propagating at the longitudinal wave speed.
For $x_1<25$ mm there are disturbances that are
pulses traveling at supersonic speeds that eventually coalesce with the
crack-like sliding region, thereby expanding it.

\subsection{Case V: A pulse-like mode followed by a crack-like mode}

\begin{figure}[tbp]
\begin{center}
\includegraphics[width=5in]{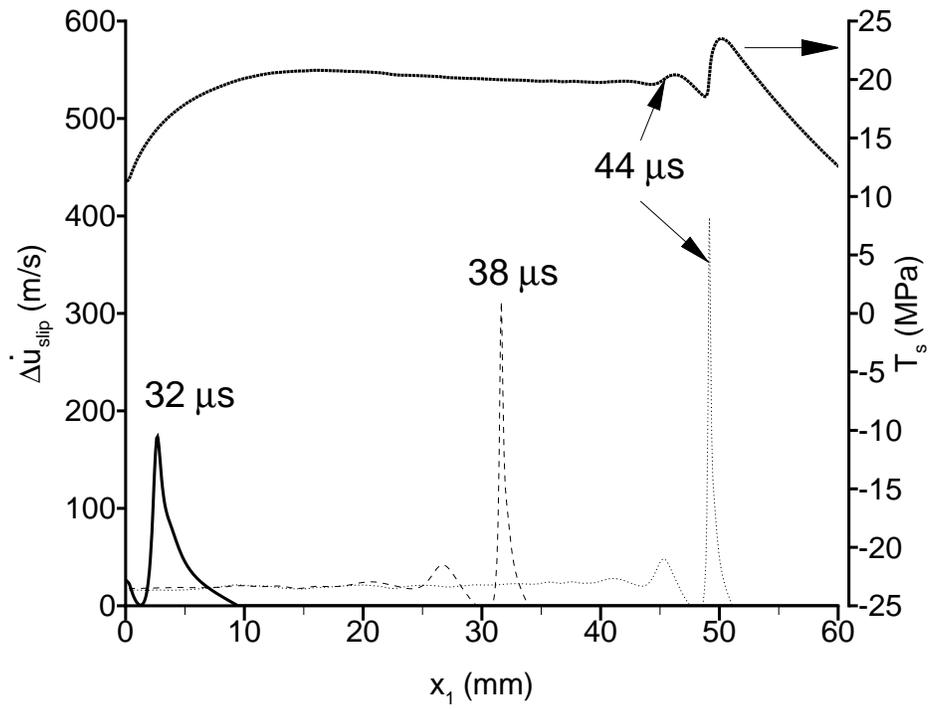}
\end{center}
\caption{Distribution of $\Delta \dot{u}_{\mathrm{slip}}$
at $t=32$ $\mu$s, $t=38$ $\mu$s and $t=44$ $\mu$s and $T_s$ at $t=44$ $\mu$s
along a portion of the interface for Case V -- $\Sigma_0=40$ MPa, 
$V_{\mathrm{imp}}=2$ m/s.}
\label{figs17}
\end{figure}

In Case V, 
$p=0.5$ and $V_1=1.0$ m/s in eqs.~(\ref{pn2}) and (\ref
{eqss}). Sliding
initiates as a slip pulse at 31 $\mu$s. Figure~\ref{figs17} shows the
distribution of $\Delta \dot{u}_{\mathrm{slip}}$ along the interface at
three times; $t=32$ $\mu$s, $t=38$ $\mu$s and $t=44.6$ $\mu$s. Also shown is
the distribution of the shear traction, $T_s$, along the interface at $t=44$ 
$\mu$s. A growing pulse is obtained with the pulse width decreasing from 8.5
mm at $t=32$ $\mu$s to 2.5 mm at $t=44$ $\mu$s and the peak value of $\Delta 
\dot{u}_{\mathrm{slip}}$ increases from 155 m/s to 392 m/s. The propagation
speed of the pulse varies slightly over the range of times in Fig.~\ref
{figs17}; 2600 m/s at $t=32$ $\mu$s and 2450 m/s at both 
$t=38$ $\mu$s and $t=44.6$ $\mu$s. After some oscillation behind the
growing pulse, $\Delta  
\dot{u}_{\mathrm{slip}}$ is constant at about $21$ m/s. The shear
traction increases to 
a maximum value of 23.5 MPa at the tip of the pulse, then drops to 18.5 MPa
in the pulse region, after which it remains almost constant at 20 MPa over
the rest of the sliding region. The apparent coefficient of friction changes
from a maximum value of 0.595 at the sliding tip to a minimum value of 0.47
at the point of maximum frictional sliding rate until it settles at a steady
state value of 0.50 along the rest of the sliding surface.

\begin{figure}[tbp]
\begin{center}
\includegraphics[width=4.0in]{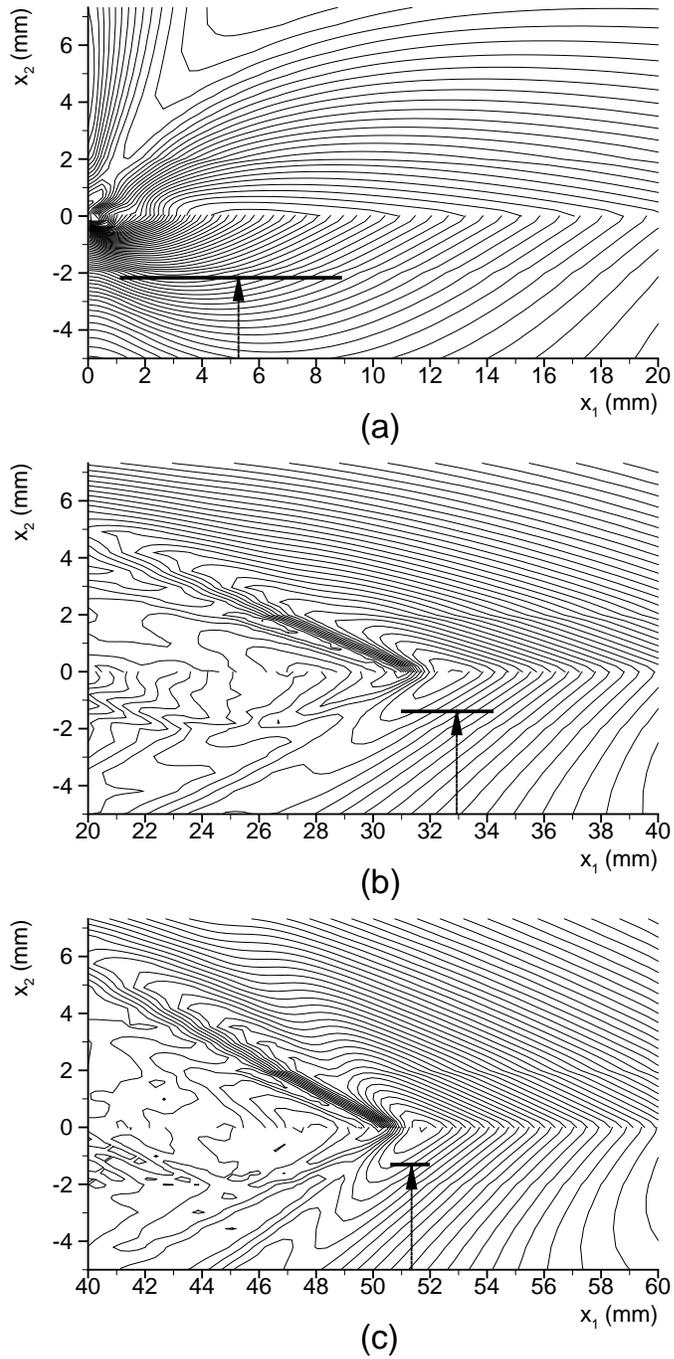}
\end{center}
\caption{Contours of $\sigma_1-\sigma_2$ (isochromatic
fringe patterns) for Case V. (a) $t=32$ $\mu$s. (b) $t=38$ $\mu$s. 
(c) $t=44.6$ $\mu$s. The dark lines indicate the location and
width of the pulses in $\Delta \dot{u}_{\mathrm{slip}}$ in 
Fig.~\ref{figs17}.}
\label{figs18}
\end{figure}

The isochromatic fringe patterns (contours of $\sigma_1-\sigma_2$) are shown
in Fig.~\ref{figs18} at three times of 32 $\mu$s, 38 $\mu$s and 44.6
$\mu$s. The arrow in 
each figure point to a line (drawn off the interface) that indicates the
width of the pulse and its position along the interface. The pulse width
decreases as the crack catches up to the pulse with time and
distance. 
At $t=38 \mu$s (Fig.~\ref{figs18}b), there are two separate Mach waves
corresponding to the slip pulse and the initiation of crack-like slip. As
the pulse width diminishes in length the tip of the crack-like region
approaches the leading tip of the pulse. From the shear Mach angles, the
propagation speed of the back of the pulse in Fig.~\ref{figs18}b is
approximately 3500 m/s, while the front of the pulse is traveling at
about 2600 m/s.

\subsection{Dependence of the sliding mode on the initial compressive stress
and impact velocity}

\begin{figure}[tbp]
\begin{center}
\includegraphics[width=6.0in]{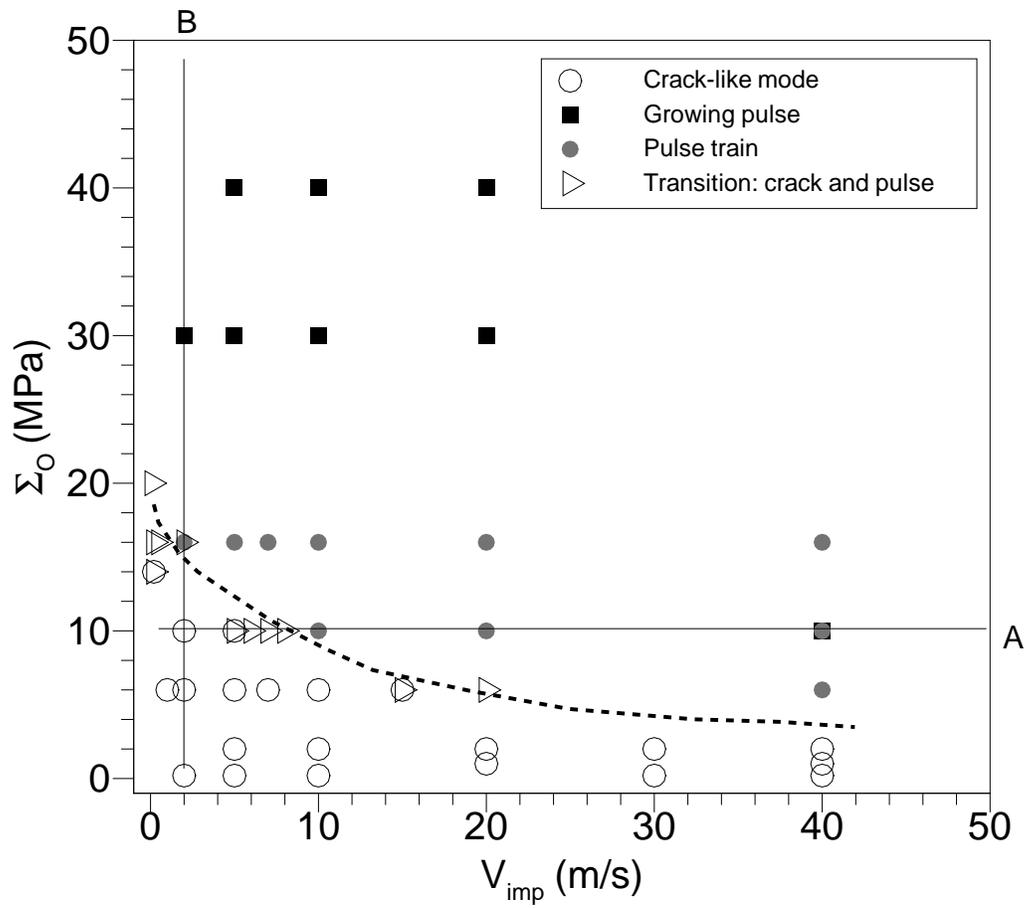}
\end{center}
\caption{Dependence of the mode of sliding on the initial
compressive stress $\Sigma_0$ and the impact velocity
$V_{\mathrm{imp}}$.
Transition cases where mixed crack-like mode and pulse-like
modes occur are shown as triangles.}
\label{fig19}
\end{figure}

The dependence of the frictional sliding mode on the initial compressive
stress, $\Sigma_0$, and the impact velocity, $V_{\mathrm{imp}}$, is shown in
Fig.~\ref{fig19}. All calculations in Fig.~\ref{fig19} were carried out
using the interface friction properties listed in Table \ref{tab:3}
and 
$C_n=0.3$ MPa/m, $C_s=0.1$ MPa/m. The dashed line indicates the boundary
between the occurrence of crack-like and pulse-like modes. It is near the
dashed line that mixed crack-like and pulse-like modes occur. Quite
generally, for low impact velocities and low values of the applied
compressive stress, the crack-like mode is obtained. On the other hand,
for a high magnitude of the initial compressive stress and a sufficiently
low impact velocity, the growing pulse mode occurs.

\begin{figure}[tbp]
\begin{center}
\includegraphics[width=3.5in]{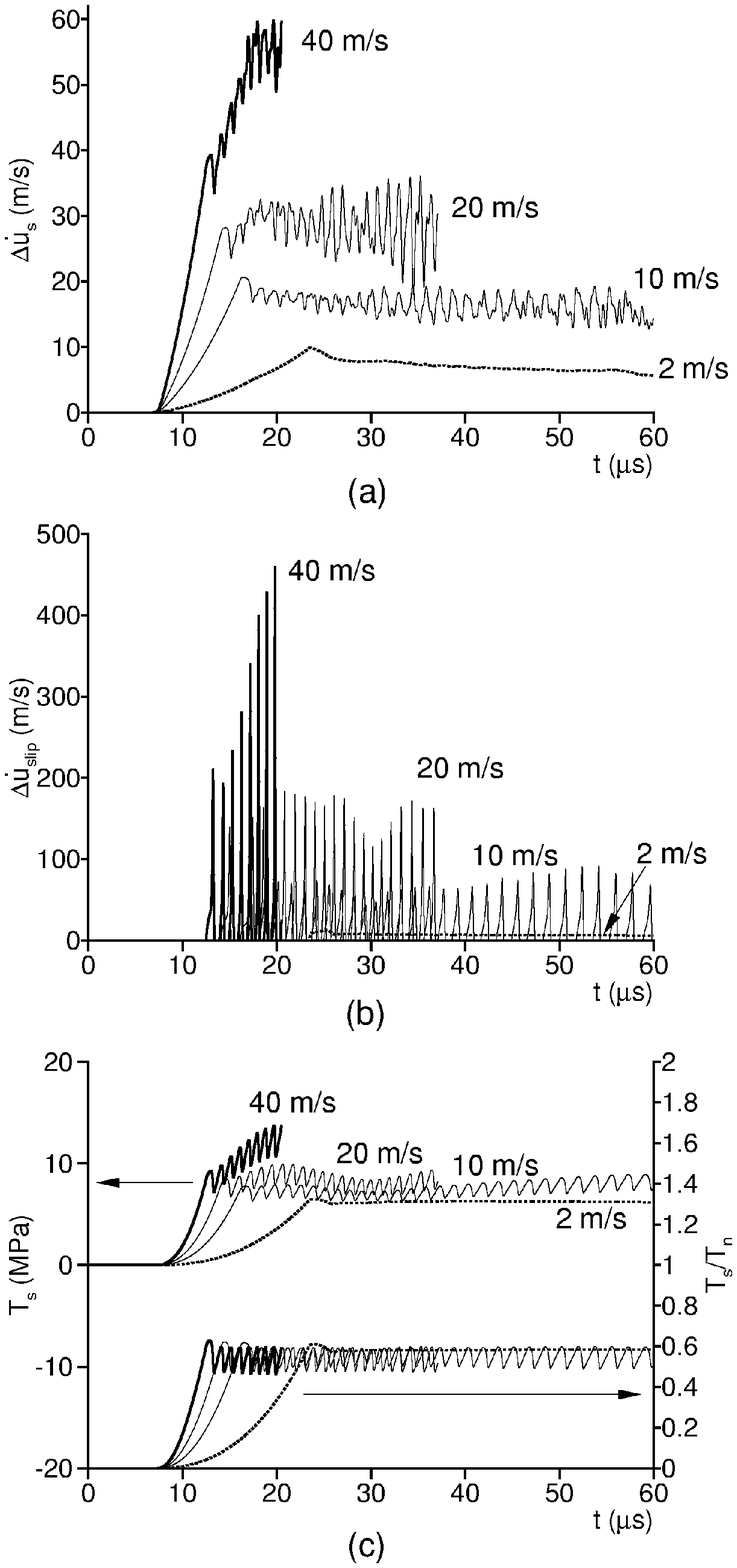}
\end{center}
\caption{Effect of varying impact velocity, 
$V_{\mathrm{imp}}$, on the frictional sliding mode for $\Sigma_0$
  fixed at $10$ MPa. (a) $\Delta\dot{u}_s$, (b)
  $\Delta\dot{u}_{\mathrm{slip}}$, (c) $T_s$ and
  $\mu_{\mathrm{app}}=T_s/T_n$.} 
\label{figs20}
\end{figure}

Figure \ref{figs20} illustrates the nature of the transition for a fixed
value of $\Sigma_0$ and varying $V_{\mathrm{imp}}$, corresponding to the
line $A$ in Fig.~\ref{fig19}. In Fig.~\ref{figs20}, $\Sigma_0$ is fixed at 10
MPa and $V_{\mathrm{imp}}$ varies between 2 m/s to 40 m/s. Values of $\Delta 
\dot{u}_s$, $\Delta \dot{u}_{\mathrm{slip}}$, $T_s$ and $\mu_{\mathrm{app}}=
T_s/T_n$ at $x_1=16.1$ mm on the interface are plotted as functions of time.
With $V_{\mathrm{imp}}=2$ m/s, essentially steady values of the velocity
jump across the interface, $\Delta \dot{u}_s$, and the frictional sliding
rate, $\Delta \dot{u}_{\mathrm{slip}}$, are attained as seen in Figs.~\ref
{figs20}a and \ref{figs20}b, respectively. With the impact velocity increased
to 10 m/s, the mode has changed to one involving a train of pulses, as seen
in Fig.~\ref{figs20}b. With a further increase in impact velocity, the pulse
amplitude increases as does the pulse frequency. Correspondingly, the
oscillations in the velocity jump across the interface increase. However, as
seen by comparing the response for $V_{\mathrm{imp}}=20$ m/s with 
$V_{\mathrm{imp}}=40$ m/s in Fig.~\ref{figs20}a, the amplitude of the
oscillations in $\Delta \dot{u}_s$ do not continue to increase with
increasing impact velocity, in contrast to the pulse amplitude in Fig.~\ref
{figs20}b. Rather, there is an increase in the `elastic' part of the velocity
jump and this is reflected in the increased magnitude of the shear
traction $T_s 
$ in Fig.~\ref{figs20}c. The oscillations in the apparent coefficient of
friction, $\mu_{\mathrm{app}}=T_s/T_n$, increase in magnitude with
increasing $V_{\mathrm{imp}}$; $\mu_{\mathrm{app}}$ changes from an almost
steady value of 0.598 at $V_{\mathrm{imp}}=2$ m/s to $\mu_{\mathrm{app}}$
oscillating between 0.598 and 0.470 at $V_{\mathrm{imp}}=40$ m/s. In
addition, the frequency of the oscillations in $\mu_{\mathrm{app}}$
increases with increasing impact velocity.

\begin{figure}[tbp]
\begin{center}
\includegraphics[width=3.5in]{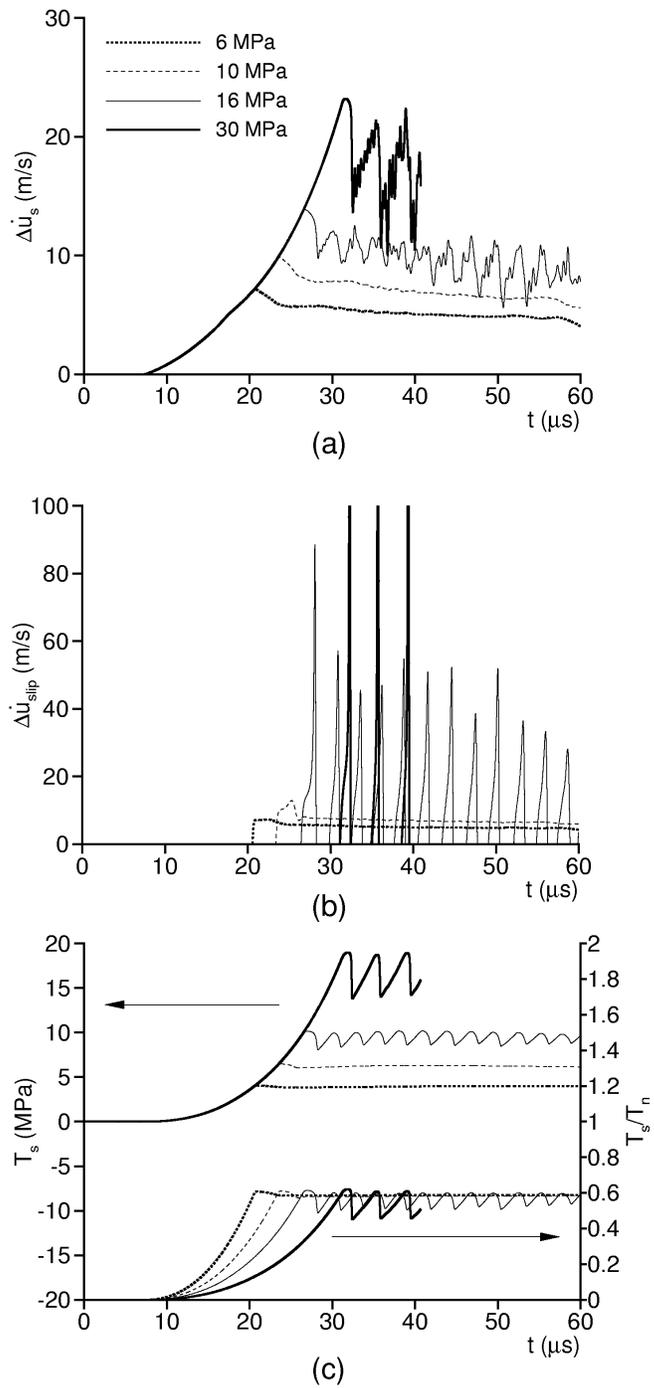}
\end{center}
\caption{Effect of varying initial compressive stress,
$\Sigma_0$, on the frictional sliding mode for $V_{\mathrm{imp}}$ fixed at 2
m/s. (a) $\Delta\dot{u}_s$, (b) $\Delta\dot{u}_{\mathrm{slip}}$, (c) $T_s$
and $\mu_{\mathrm{app}}=T_s/T_n$.}
\label{figs21}
\end{figure}

The nature of the transition for $V_{\mathrm{imp}}$ fixed at 2 m/s and with
varying $\Sigma_0$, the line $B$ in Fig.~\ref{fig19}, is illustrated
in Fig.~\ref{figs21}. As in Fig.~\ref{figs20}, the evolution with time
of $\Delta \dot{u}_s$, $\Delta \dot{u}_{\mathrm{slip}}$, $T_s$ and
$\mu_{\mathrm{app}}$ at a 
fixed point along the interface ($x_1=16.1$ mm) is shown. With increasing
initial compressive stress, frictional sliding initiates later (Figs.~\ref
{figs21}a and \ref{figs21}b). For both $\Sigma_0=6$ MPa and $\Sigma_0=10$ MPa,
crack-like behavior is obtained. The values of $\Delta \dot{u}_s$ and 
$\Delta \dot{u}_{\mathrm{slip}}$ at the beginning of frictional sliding
exceed their subsequent steady-state values, which is a consequence of
the
direct effect in the rate- and state-dependent friction relation. With 
$\Sigma_0=16$ MPa, a train of pulses in frictional sliding rate (Fig.~\ref
{figs21}b) occurs, with oscillations both in $\Delta \dot{u}_s$ (Fig.~\ref
{figs21}a) and in $T_s$ (Fig.~\ref{figs21}c). When the compressive load is
increased to $\Sigma_0=30$ MPa, steeper 
and narrower pulses are obtained, Fig.~\ref
{figs21}b, and the peak value of $\Delta \dot{u}_{\mathrm{slip}}$
reaches $\approx 820$ m/s, much exceeding the scale in
Fig.~\ref{figs21}b. These slip pulses increase in amplitude
as they propagate and the peak $\Delta \dot{u}_{\mathrm{slip}}$ can be
of the order of $c_s$.
For a sufficiently high value of $\Sigma_0$, 
the closely trailing slip pulses are suppressed. Then, a
trailing slip pulse only occurs after some distance and time has passed so
that the pulses are well-separated. 
Such growing pulses have been obtained in studies of
sliding between dissimilar elastic solids using rate- and state-dependent
friction laws (Ben-Zion and Huang, 2002; Adda-Bedia and Ben Amar,
2003). It is worth noting that if the scale of observation is 
sufficiently small compared with the pulse spacing, only a single pulse is
observed which resembles the slip pulses discussed in relation to
earthquakes (Heaton, 1990).

The nature of the slip pulses that
occur when $\Sigma _{0}$ is increased (line B in Fig.~\ref{fig19})
differ from those obtained when $V_{\mathrm{imp}}$ is increased (line
A in Fig.~\ref{fig19}). With fixed $\Sigma _{0}$,  increasing
$V_{\mathrm{imp}}$ increases the magnitude of the slip pulses while
the pulse spacing and the 
pulse width decrease. For $\Sigma _{0}=16$ MPa, it was found that the pulse
width to pulse spacing ratio remained constant with varying 
$V_{\mathrm{imp}}$ and this ratio was equal to one-half. This behavior is
analogous to the oscillatory stick-slip behavior observed by Baumberger et
al. (1994). The extent to which the pulse width to pulse spacing ratio
depends on $\Sigma _{0}$ remains to be determined. In both
Figs.~\ref{figs20}c and \ref{figs21}c, the peak values of the apparent
coefficient of friction do not increase with increasing
$V_{\mathrm{imp}}$ (Fig.~\ref{figs20}c) or $\Sigma_0$
(Fig.~\ref{figs21}c) even though the magnitude of $T_s$ does, thus
implying that there is a corresponding increase in $T_n$.

To illustrate the evolution of sliding in a transition mode, we 
describe this evolution for the
calculation with $\Sigma_0=10$ MPa, $V_{\mathrm{imp}}=7$ m/s (marked by
one of the triangles in Fig.~\ref{fig19}). Oscillations occur
in $\Delta \dot{u}_{\mathrm{slip}}$ along the interface, with the highest
peak near the sliding tip and decreasing in amplitude behind the tip. When
sliding has progressed about 50 mm along the interface, the amplitude of the
first three oscillations is large enough for $\Delta \dot{u}_{\mathrm{slip}}$
to vanish at the crests leading to sticking regions (however the maximum
value of $\Delta \dot{u}_{\mathrm{slip}}$ generally remains a small fraction
of an elastic wave speed). The amplitude of the following oscillations is
much smaller and sticking does not occur. Thus, we obtain a crack-like
sliding region following three slip-pulses of decreasing magnitude.

Other loading combinations of $\Sigma _{0}$ and $V_{\mathrm{imp}}$ on the
transition line in Fig.~\ref{fig19}, lead to similar oscillations in the
slip rate or to unsteady behaviors, such as mixed crack-like and pulse-like
sliding modes that evolve with both time and distance. This behavior
is reminiscent of
that found in the analysis of spring-slider system obeying a rate- and
state-dependent friction law (Ruina, 1983; Gu et al., 1984) where it was
shown that the transition from steady sliding to stick-slip goes through
several oscillations, period doubling and self-sustained periodic
oscillations. Gu and Wong (1991,1994) also found this transition to be
dependent on the load point driving velocity in addition to the stiffness
and constitutive parameters.

\section{Discussion}

The experiments and
calculations for symmetric loading with no sliding, permit the
features of the response associated with sliding to 
be identified. Also, the good agreement between the experiments and the
calculations for symmetric loading provides confidence that the wave
propagation aspects of the 
experiments are well-represented in the calculations. Features that are
common between the symmetric loading experiments and the corresponding
calculations include the cusp in the photoelastic fringes at the interface,
the formation of a head wave emanating from the loading wave front and the
distortions in the photoelastic stress field behind the loading wave due to
the trapezoidal nature of the input.

A guideline for assessing mesh resolution (Zheng and Rice, 1998) is
that the mesh spacing ($0.3$ mm along the interface in the uniform
mesh region) should be significantly less than a characteristic
length, $h^*$, associated with the friction law,
\[
h^*=\vert 2
G L_0/[\pi \Delta \dot{u}_{\mathrm{slip}} (d\tau_{ss}/d \Delta
\dot{u}_{\mathrm{slip}})] \vert
\]
where $G$ is the shear modulus and
$\tau_{ss}$ is the shear stress under steady state sliding. For a
normal stress of $10$ MPa, $h^*$ attains its minimum of $0.81$ mm for
$\Delta \dot{u}_{\mathrm{slip}}=V_1=26$ m/s. Using $\Delta
\dot{u}_{\mathrm{slip}}=V_1=7$ m/s as a representative value for Case
I, $h^*=1.8$ mm. For Case IV, where the normal stress is $\approx 1$
MPa and $\Delta \dot{u}_{\mathrm{slip}}$ ranges from about 10 m/s to
100 m/s, the minimum value of $h^*$ is $8.1$ mm and for much of the
velocity range is significantly greater than that. It is worth noting
that it is for Case IV that a supersonic pulse is obtained. For the
other cases, $h^*$ typically has values between those for Cases I and
IV.

Several calculations were carried out to assess the mesh dependence of
the results. For Case I (crack-like mode) and for Case III (pulse-like
mode), calculations were carried out with the mesh spacing halved
along the interface (0.15 mm in the uniform region) so that the number
of elements was doubled. For Case I, with the finer mesh the
transition region at the sliding tip consisted of four elements
(sixteen integration points) as opposed to two elements (eight
integration points) with the coarser mesh. The values of $\Delta
\dot{u}_{\mathrm{slip}}$ obtained from the two calculations nearly
coincide. The mean values of the curves of $V_{\rm tip}$ versus time
computed using the finer mesh also essentially coincide with the
curves in Fig.~\ref{figs6} but the oscillations about the mean differ
somewhat. In addition, the effect of mesh size was further explored
for a crack-like mode and a pulse like mode using frictional
parameters that differ from those used in Cases I and II.  The
reference mesh shown in Fig.~\ref{fig1}b and two other meshes were
used for this. All the meshes have the smallest elements near the
interface and the impact edge as for the reference mesh in
Fig.~\ref{fig1}b. The coarse mesh consisted of $1,152$ elements and
$4,812$ degrees of freedom, with the smallest element size $2.1$ mm
$\times $ $1.0$ mm.  The fine mesh consisted of $56,832$ elements and
$229,364$ degrees of freedom, with the smallest element size $0.075$
mm $ \times $ $0.1$ mm.  The slip rate at a point along the interface
was plotted as a function of time for both meshes. In the calculation
with the crack-like sliding mode, the results from all three
calculations are in very good agreement for the propagation speed
except that increased fluctuations around a steady slip rate occur
with the coarse mesh. For the pulse-like mode calculation, there is
only a small difference between the calculations using the three
meshes for the predicted magnitude and timing of the pulses.

The values used for the interface stiffnesses $C_n$ and $C_s$ in eqs.~(\ref
{eqtime11}) and (\ref{eqtime1}) play a significant role both for symmetric
loading (with no sliding) and for asymmetric loading (with frictional
sliding). For symmetric loading, Fig.~\ref{fig7}, the width of the head wave
emanating from the loading wave front decreases with increasing interface
stiffness. Because in the experiments the width of this head wave expands
with decreasing applied load, the symmetric loading geometry can be used to
calibrate the stiffness values. The values used in 
most of the calculations,
$C_n=0.3$ MPa/m and $C_s=0.1$ MPa/m, give rather good agreement for
symmetrically loaded specimens with $\Sigma_0$ in the range $5-40$ MPa.

To assess the dependence of the sliding mode on the interface stiffness
values employed, asymmetric loading calculations were carried out using
various values of $C_{n}$ and $C_{s}$. For given values of $\Sigma _{0}$ and 
$V_{\mathrm{imp}}$, the frictional sliding mode that occurs depends on 
$C_{n}$ and $C_{s}$. With $C_{n}=3.0$ MPa/m and $C_{s}=1.0$ MPa/m,
only crack-like 
behavior was obtained for the range of values of $\Sigma _{0}$ and 
$V_{\mathrm{imp}}$ used in Fig.~\ref{fig19}. On the other hand, 
various calculations were carried out with $C_{n}=0.03$ 
MPa/m and $C_{s}=0.01$ MPa/m and a train of pulses was not
obtained, although as shown for Case IV
pulses coalescing to form a crack-like
sliding mode did occur. Also, pulse steepening became more pronounced
when $C_{n}$ and $C_{s}$  were decreased.

The elasticity of the interface plays a fundamental role 
because the frictional slip 
$\Delta \dot{u}_{\mathrm{slip}}$ is an internal variable determined by
the frictional 
law (analogous to the plastic strain rate in plasticity theory). The
observable quantity is the jump in displacement rate across the
interface, 
$\Delta \dot{u}_{\mathrm{s}}$. The pulses occur in $\Delta
\dot{u}_{\mathrm{slip}}$ not in $\Delta \dot{u}_{\mathrm{s}}$ and,
hence, according to our 
numerical results, are not directly observable in terms of a velocity jump
distribution along an interface. What is observable is a pulse-type
traction rate distribution (Figs.~\ref{figs10}b and \ref{figs12}b). At a given
spatial location, the oscillatory nature of the pulse-like modes is seen in
the time evolution of $\Delta \dot{u}_{\mathrm{s}}$
(Fig.~\ref{figs20}a).

Calculations modeling dynamic slip in earthquakes, e.g.
Cochard and Madariaga (1994, 1996), Perrin et al. (1995), Beeler and Tullis
(1996) (see also Ben-Zion, 2001; Rice, 2001; Nielsen and Madariaga,
2003), exhibit 
both crack-like and pulse-like sliding modes. In analyzing 
conditions that set the 
mode of sliding, Zheng and Rice (1998) considered the special case of
identical material properties on both sides of the slip surface.
Specifically, Zheng and Rice (1998) considered elastic half-spaces that are
subject to a uniform shear stress $\tau _{0}^{b}$ outside a perturbed region
where slip nucleates. The criterion delineating between crack-like and
pulse-like sliding involves the magnitude of $\tau _{0}^{b}$, the
elastic moduli and shear wave speed of the half-spaces, the sliding speed
and the sliding speed dependence of the interface friction law. The
criterion of Zheng and Rice (1998) predicts that low values of 
$\tau_{0}^{b}$ favor pulse-like sliding and higher values crack-like
sliding.

The circumstances considered here are quite different from those in Zheng
and Rice (1998). In particular, it is important to note that the parameters
in Fig.~\ref{fig19} are the initial \textit{compressive} stress and the 
\textit{impact} velocity. In this parameter space, a clear demarcation
is found between 
the regime where the crack-like and pulse-like modes occur; a low initial
compressive stress and/or a low impact velocity favor the crack-like mode.

The experiments and calculations exhibit a variety of common features
including: 
\begin{description}
\item{(i)} Direct calculation from the time history of the
position of the sliding tip and Mach lines in the contours of
$\sigma _{1}-\sigma _{2}$  indicate that the sliding tip 
travels at a speed greater than $\sqrt{2} c_s$ and that a
trailing pulse travels faster than $c_l$.

\item{(ii)} The fringe patterns
in front of the sliding region have similar characteristic shapes. 

\item{(iii)} 
Crack-like sliding occurs at low impact velocities
and a sliding mode with a pulse is obtained at higher impact
velocities.

\item{(iv)}
Two shock waves eventually merge into a single shock wave as the trailing
shock catches up to the leading one. The tip of the trailing pulse
travels at a speed exceeding $c_l$.
\end{description}

The calculations exhibit two sliding modes with two approaching
shock waves. One involves trailing pulses
traveling faster than the leading pulse that eventually coalesce to form one
sliding region (Figs. \ref{fig15} and \ref{fig16}). The other involves a
steepening and narrowing pulse followed by a crack-like sliding region
(Figs. \ref{figs17} and \ref{figs18}). 

There is some evidence in other contexts regarding the range of sliding
modes that emerge from our calculations, more specifically regarding
the pulse and pulse train modes of frictional sliding. Rubio and
Galiano (1994) carried out 
experiments on sheared gels sliding along smooth glass surfaces and observed
sliding via propagation of a quasi-periodic pattern of sliding zones of
finite width separated by non-moving regions, having a propagation speed of
the order of $c_{l}$ of the gel. This is consistent with our observations of
relative speeds of pulse propagation that are three order of magnitude
higher than the imposed impact velocities. The experiments of Rubio and
Galiano (1994) motivated an analysis by Caroli (2000) of sliding of a
viscoelastic solid on a rigid substrate with a rate and state friction law
that showed a pulse train. Sliding between a rectangular Polyurethane
slab and a compressed Araldite plate was studied experimentally by
Mouwakeh et al. (1991). The velocity measured at a fixed spatial point was
strongly oscillatory. Also, the measured traction exhibited oscillations.
The traction versus time curve in Mouwakeh et al. (1991) resembles one of
the oscillatory curve in Fig.~\ref{figs21}c. The velocity versus time curve
is qualitatively similar to the intermediate curves in Fig.~\ref{figs21}b.
However it is unclear whether $\Delta \dot{u}_{\mathrm{slip}}$ or $\Delta 
\dot{u}_{s}$ is plotted in Mouwakeh et al. (1991, Fig. 5a).

Fiber pull-out experiments exhibit a variety of complex
frictional phenomena, e.g. Tsai and Kim (1996) and Li et al. (2002).
In particular, Tsai and Kim (1996) observed three frictional
sliding modes: steady sliding of the entire contact surface, stick-slip
sliding of the entire contact surface, and sliding through the generation of
concentrated sliding pulses. In one set of experiments, with sliding along
the entire contact surface, the pull-out force was steady for a low
interface pressure and was oscillatory for a high interface pressure,
analogous to the behavior in our calculations where a crack-like sliding
mode occurs for low values of $\Sigma_0$ and a pulse-like mode for high
values of $\Sigma_0$. However, 
Tsai and Kim (1996) found that the oscillation amplitude decreased with
increasing pull-out speed. This trend, also found in Baumberger et al.
(1994) and Povirk and Needleman (1993), under quasi-static loading
conditions, is opposite to what occurs in our analyses under dynamic
loading conditions.

Brune et al. (1993) and Anooshehpoor and Brune (1994) observed two modes of
sliding along the interface between identical foam rubber blocks; one
involving wrinkle-like waves and the other a crack-like sliding mode.
Characteristics of the wrinkle-like wave mode in these experiments are
similar to those obtained for slip-pulses in our calculations; for example,
an increasing normal component of particle motion with increasing normal
stress which is analogous to the slip pulse amplitude increasing with 
$\Sigma_0$ in our calculations. Interestingly, it was found that the
frictional heating did not change with shear stress in the wrinkle-like wave
mode whereas it increased linearly with shear stress in the crack-like
mode (see also Andrews and Ben-Zion, 1997, Fig. 12).
In experiments on sliding of two dissimilar blocks of foam rubber,
Anooshehpoor and Brune (1999) observed that sliding occurred through a
series of multiple opening pulses. They also found that the sharpness and
the amplitude of the pulses either increased or decreased with time and was
not steady. Such opening pulses, often referred to as Schallamach
(1971) waves, have been observed experimentally in various bimaterial
systems (e.g. Schallamach, 1971; Anooshehpoor and Brune, 1999) and in
simulations (Coker et al., 2003). There is evidence for slip pulses,
on which attention is focused here, in the experiments of Mouwakeh et
al. (1991) and Tsai and Kim (1996).

Figure \ref{figs14}  shows that the
displacement distribution from a pulse train slip mode and from a crack-like
slip mode are quite similar. This might be the reason that there is ample
experimental evidence of slip pulses in soft materials such as gels, whereas
there is no conclusive experimental evidence for slip pulses in engineering
materials where the displacement magnitudes are very small compared to those
in soft materials. However, the shear traction distribution along the
interface in Fig.~\ref{figs14} does provide an indicator of the slip-mode.
The variation of the shear resistance and the step increase in displacement
in Fig.~\ref{figs14} is similar to the response seen in rock friction
experiments (Tullis, 1996) and in fiber pull-out experiments (Tsai and Kim,
1996). Since the resulting slip from a train of pulses is hard to
distinguish from that occurring with the crack-like slip mode, our results
suggest the possibility that slip pulses may be more common than previously
appreciated. It remains to be determined whether there is a significant
difference between the heat generation associated with the pulse-train
mode and that associated with the crack-like mode.

\section{Conclusions}

Frictional sliding between two Homalite plates subject to a compressive
stress was investigated both experimentally and computationally. Sliding
under shear loading was induced by impacting one plate parallel to the
interface. The experiments yield information on the sliding speed and the
stress state in the material using dynamic photoelasticity. A plane stress
initial/boundary value problem is analyzed using a framework where
sliding is modeled by a rate- and state-dependent frictional interface
constitutive relation. 

\begin{itemize}
\item  The experiments exhibit both pulse-like and crack-like modes of
sliding.

\item  In the experiments,
the speed of the leading pulse or the leading edge of the
crack-like sliding region ranges from somewhat above $\sqrt{2}c_s$ to $c_l$.
This propagation speed increases with increasing impact velocity and
decreasing compressive stress. A speed exceeding $c_l$ was seen for a
trailing pulse.

\item  A variety of frictional sliding modes were obtained in the 
calculations,  depending on the initial compressive stress, the impact
velocity and  the interface characterization: a crack-like mode; a 
pulse-like mode with well-separated pulses that increase in  amplitude; and
a train of pulses that propagate with an  essentially constant amplitude. In
addition, combinations of these  modes occurred as well as transitions
between modes. This variety of sliding modes emerges even though there is no
elastic mismatch across the interface.

\item The slip resulting from the pulse-train mode and that resulting from
the crack-like mode are hard to distinguish.

\item  In all calculations the speed of the leading pulse or the
leading edge of the crack-like sliding region exceeds $\sqrt{2}c_s$ and is 
close to (or slightly exceeds) $c_l$. As in the experiments, trailing
pulses with a speed exceeding $c_l$ are found.

\item  The elasticity of the interface plays a significant role in setting
the mode of sliding.

\item  The range of sliding modes obtained appear to be generic, arising
in a wide variety of configurations and applications, and at a wide variety
of size scales.
\end{itemize}

\section*{Acknowledgments}

\small{

\noindent DC and AN are pleased to acknowledge support from the Office of
Naval support from the Research through grant N00014-97-1-0179 and from the
General Motors Cooperative Research Laboratory at Brown University. GL and
AJR are grateful for support from the Office of Naval Research through grant
N00014-02-1-0522. We are indebted to Professor J.R. Rice of Harvard
University and to Professors K.S. Kim and  T.E. Tullis of Brown
University for stimulating 
comments and insightful suggestions during the course of this work.
}

\section*{References}

\small{

\vspace{3pt} \noindent Abraham, F.F., 2001. The atomic dynamics of
fracture. J. Mech. Phys. Solids, 49, 2095-2111.

\vspace{3pt} \noindent Abraham, F.F., Gao, H.J., 2000. How fast can
cracks propagate? Phys. Rev. Lett., 84, 3113-3116.

\vspace{3pt} \noindent Adams, G.G., 1995. Self-excited oscillations of two
elastic half-spaces sliding with a constant coefficient of friction. J.
Appl. Mech., 62, 867-872.

\vspace{3pt} \noindent Adda-Bedia, M., Ben Amar, M., 2003. Self-
sustained slip pulses of finite size between dissimilar materials.
J. Mech. Phys. Solids, 51, 1849-1861.

\vspace{3pt} \noindent Andrews, D.J., 1976. Rupture velocity of plain
strain shear cracks. J. Geophys. Res., 81, 5679-5687.

\vspace{3pt} \noindent Andrews, D.J., Ben-Zion, Y., 1997. Wrinkle-like
slip pulse on a fault between different materials. J. Geophys. Res., 102,
553-571.

\vspace{3pt} \noindent Anooshehpoor, A., Brune, J.N.,1994.  Frictional
heat-generation and seismic radiation in a foam rubber model of
earthquakes. Pure Appl. Geophys., 142, 735-747.

\vspace{3pt} \noindent Anooshehpoor, A., Brune, J.N., 1999.  
Wrinkle-like Weertman pulse at the interface between two blocks of foam
rubber with different velocities. Geophys. Res. Letts., 26, 2025-2028.

\vspace{3pt} \noindent Archuleta, R.J., 1984. A faulting model for the 1979
Imperial Valley earthquake. J. Geophys. Res., 89, 4559-4585.

\vspace{3pt} \noindent Barquins, M., Courtel, R. 1975. Rubber friction
and rheology of viscoelastic contact. Wear, 32, 133-150.

\vspace{3pt} \noindent Baumberger, T., Heslot, F.,  Perrin, B., 1994.
Crossover from creep to inertial motion in friction dynamics. Nature, 367,
544-547.

\vspace{3pt} \noindent Beeler, N.M., Tullis, T.E., 1996.  Self-healing
slip pulses in dynamic rupture models due to velocity-dependent
strength. Bull. Seismol. Soc. Amer., 86, 1130-1148.

\vspace{3pt} \noindent Belytschko, T., Chiapetta, R.L., Bartel, H.D.,
1976. Efficient large scale non-linear transient analysis by finite
elements. Int. J. Numer. Meth. Engr., 10, 579-596.

\noindent Ben-Zion, Y., 2001. Dynamic ruptures in recent models of 
earthquake faults. J. Mech. Phys. Solids, 49, 2209-2244.

\vspace{3pt} \noindent Ben-Zion, Y., Andrews, D.J., 1998. Properties and
implications of dynamic rupture along a material interface. Bull. Seismol.
Soc. Amer., 88, 1085-1094.

\vspace{3pt} \noindent Ben-Zion, Y., Huang, Y., 2002. Dynamic rupture on
an interface between a compliant fault zone layer and a stiffer surrounding
solid. J. Geophys. Res., 107, B2 article no 2042. 

\vspace{3pt} \noindent Ben-Zion, Y., Rice, J. R. 1997. Dynamics
simulations 
of slip on a smooth fault in an elastic solid. J. Geophys. Res., 102,
17,771-17,1784.

\vspace{3pt} \noindent Bizzarri, A., Cocco, M., Andrews, D.J., Boschi,
E., 2001. Solving the dynamic rupture problem with different numerical
approaches and constitutive laws. Geophys. J. Int., 144, 656-678.

\vspace{3pt} \noindent Bouchon, M., Bouin, M.P., Karabulut, H., Toks\"oz,
M.N., Dietrich, M., Rosakis, A.J., 2001. How fast is rupture during
an earthquake? New sights from the 1999 Turkey earthquakes. Geophys. Res.
Lett., 28, 2723-2726.

\vspace{3pt} \noindent Bouchon, M., Vallee, M., 2003. Observation of
long supershear rupture during the magnitude 8.1 Kunlunshan earthquake.
Science, 301, 824-826.

\vspace{3pt} \noindent Broberg, K.B., 1995. Intersonic mode II crack
expansion. Arch. Mech., 47, 859-871.

\vspace{3pt} \noindent Broberg, K.B., 1996. How fast can a crack go? Mater.
Sci., 32, 80-86.

\vspace{3pt} \noindent Broberg, K.B., 1999. {\it Cracks and fracture}. 
Academic Press, London.

\vspace{3pt} \noindent Brune, J.N., Brown, S., Johnson P.A., 1993.
Rupture mechanism and interface separation in foam rubber models of
earthquakes - a possible solution to the heat-flow paradox and the
paradox of large overthrusts.  Tectonophys., 218, 59-67.

\vspace{3pt} \noindent Caroli, C., 2000. Slip pulses at a sheared frictional
viscoelastic/nondeformable interface. Phys. Rew. E 62, 1729-1737.

\vspace{3pt} \noindent Cochard, A., Madariaga, R., 1994.  Dynamic
Faulting under Rate-Dependent Friction. Pure Appl. Geophys., 142,
419-445. 

\vspace{3pt} \noindent Cochard, A., Madariaga, R.,
1996. Complexity of 
seismicity due to highly rate-dependent friction. J.
Geophys. Res. Solid Earth, 101, 25321-25336.

\vspace{3pt} \noindent Cochard, A., Rice, J.R., 2000. Fault
rupture between 
dissimilar materials: Ill-posedness, regularization, and slip-pulse
response. J. Geophys. Res., 105, 25891-25907.

\vspace{3pt} \noindent Coker, D., Rosakis, A.J., 2001. Experimental
observations of intersonic crack growth in asymmetrically loaded
unidirectional composite plates. Phil. Mag. A, 81, 571-595.

\vspace{3pt} \noindent Coker, D., Rosakis, A.J., Needleman, A., 2003. 
Dynamic crack growth along a polymer composite-homalite interface. 
J. Mech. Phys. Solids, 51, 425-460.

\vspace{3pt} \noindent Dieterich, J.H., 1979. Modeling of rock  friction 1.
Experimental results and constitutive equations.  J. Geophys. Res. 84,
2161-2168.

\vspace{3pt} \noindent Dunham, E. M., Favreau, P., Carlson, J.M., 2003.
A supershear transition mechanism for cracks. Science, 299, 1557-1559.

\vspace{3pt} \noindent Freund, L.B., 1998. {\it Dynamic fracture
mechanics}. Cambridge University Press, Cambridge, UK.

\vspace{3pt} \noindent Gao, H.J., Huang, Y., Abraham, F.F., 2001.
Continuum and atomistic studies of intersonic crack propagation. J. Mech.
Phys. Solids, 49, 2113-2132.

\vspace{3pt} \noindent Gu, J.-C., Rice, J.R., Ruina, A.L.,
Tse, S. T., 1984.  Slip motion and stability of a single degree of
freedom elastic system with rate and state dependent friction.  J.
Mech. Phys. Solids 32, 167-196.

\vspace{3pt} \noindent Gu, Y.J., Wong, T.F., 1991.  
Effects of loading velocity,
stiffness, and inertia on the dynamics of a single degree of freedom
spring-slider system. J. Geophys. Res., 96, 21677-21691.

\vspace{3pt} \noindent
Gu, Y.J., Wong, T.F., 1994.  Development of shear localization in
simulated quartz gouge - effect of cumulative slip and gouge
particle-size. Pure Appl. Geophys., 143, 387-423.

\vspace{3pt} \noindent Hao, S., Wing, K.L., Klein, P.A., Rosakis, A.J., 
2004. Modeling and simulation of intersonic crack growth. Int. J. Solids
Struct., 41, 1773-1799.

\vspace{3pt} \noindent Heaton, T.H., 1990. Evidence for and implications of
self-healing pulses of slip in earthquake rupture. Phys. Earth
Planet In., 64, 1-20.

\vspace{3pt} \noindent Hernandez, B., Cotton, F., Campillo, M., 1999.
Contribution of Radar interferometry to a two-step inversion of the
kinematic process of the 1992 Landers earthquake. J. Geophys. Res., 104,
13083-13099.

\vspace{3pt} \noindent Krieg, R.O., Key, S.W., 1973.  
Transient shell response
by numerical time integration. Int. J. Numer.  Meths. Engrg., 7,
273-286.

\vspace{3pt} \noindent Li, Z., Bi, X., Lambros, J., Geubelle, P.H.,
2002. Dynamic fiber debonding and frictional push-out in 
model composite systems: Experimental observations. Exp. Mech., 42,
417-425.

\vspace{3pt} \noindent Lin, A., Guo, J., Zeng, Q., Dang, G., He, W., 
Zhao, Y., 2002. Co-Seismic strike-slip and rupture length produced by the
2001 Ms 8.1 central Kunlun earthquake. Science, 296, 2015-2017.

\vspace{3pt} \noindent Linker, M.F., Dieterich, J.H., 1992. Effects of
variable normal stress on rock friction: Observations and constitutive
equations. J. Geophys. Res. 97, 4923-4940.

\vspace{3pt} \noindent Lykotrafitis, G., Coker, D., Rosakis, A.J.,
Needleman, A., 2004. in preparation. 

\vspace{3pt} \noindent Madariaga, R., Olsen, K.B., 2000. Criticality of
rupture dynamics in 3-D. Pure Appl. Geophys., 157, 1981-2001.

\vspace{3pt} \noindent Mouwakeh, M., Villechaise, B., Godet, M., 1991.
Quantitative study of interface sliding phenomena in a two-body contact.
Eur. J. Mech. A/Solids 10, 545-555.

\vspace{3pt} \noindent Needleman, A., 1999. An analysis of intersonic crack
growth under shear loading. J. Appl. Mech., 66, 847-857.

\vspace{3pt} \noindent Nielsen, S., Madariaga, R., 2003. On the
self-healing fracture mode. Bull. Seismol. Soc. Amer., 93, 2375-2388.

\vspace{3pt} \noindent Olsen, K.B., Madariaga, R., Archuleta, R. J.,
1997. Three-dimensional dynamic simulation of the 1992 Landers earthquake.
Science, 278, 834-838.

\vspace{3pt} \noindent Peirce, D., Shih, C.F., Needleman, A., 1984. A
tangent modulus method for rate dependent solids. Comp. Struct., 18, 875-887.

\vspace{3pt} \noindent Perrin, G., Rice, J.R., Zheng, G., 1995.
Self-healing slip pulse on a frictional surface. J. Mech. Phys. Solids, 43,
1461-1495.

\vspace{3pt} \noindent Povirk, G.L., Needleman, A., 1993. Finite element
simulations of fiber pull-out. J. Eng. Mat. Tech. 115, 286-291.

\vspace{3pt} \noindent Prakash, V., Clifton, R.J., 1993. Pressure-shear
plate impact measurement of dynamic friction for high speed machining
applications. Proc. 7th Int. Congress on Exp. Mech., Society of Experimental
Mechanics, Bethel, CT, pp. 556-564.

\vspace{3pt} \noindent Prakash, V., 1998. Frictional response of sliding
interfaces subjected to time varying normal pressures. J. Tribol.,
120, 97-102.

\vspace{3pt} \noindent Ranjith, K., Rice, J.R., 2001. Slip dynamics at an
interface between dissimilar materials, J. Mech. Phys. Solids, 49, 341-361.

\vspace{3pt} \noindent Renardy, M., 1992. Ill-posedness at the boundary for
elastic solids sliding under Coulomb friction. J. Elast., 27, 281-287.

\vspace{3pt} \noindent Rice, J.R., 2001. New perspectives on crack and
fault dynamics. In: H. Aref and J. W. Philips (Eds.), Mechanics for a New
Millennium. Kluwer Academic Publishers, Dordrecht, 1-24.

\vspace{3pt} \noindent Rice, J.R., Ben-Zion, Y., 1996. Slip
complexity in earthquake fault models. Proc. Nat. Acad. Sci. USA, 93,
3811-3818.

\vspace{3pt} \noindent Rice, J.R., Lapusta, N., Ranjith, K., 2001. Rate-
and state-dependent friction and the stability of sliding between
elastically deformable solids, J. Mech. Phys. Solids, 49, 1865-1898.

\vspace{3pt} \noindent Rice, J.R., Ruina, A.L., 1983. Stability of
frictional sliding. J. Appl. Mech., 50, 343-349.

\vspace{3pt} \noindent Rosakis, A.J., 2002. Intersonic shear cracks and
fault ruptures. Adv. Phys., 51, No. 4, 1189-1257.

\vspace{3pt} \noindent Rosakis, A.J., Samudrala, O., Coker, D., 1999.
Cracks faster than the shear wave speed. Science, 284, 1337-1340.

\vspace{3pt} \noindent Rosakis, A.J., Samudrala, O., Coker, D., 2000.
Intersonic shear crack growth along weak planes. Mat. Res. Innov., 3,
236-243. 

\vspace{3pt} \noindent Rubio, M.A., Galeano, J., 1994. Stick-slip
dynamics in the relaxation of stresses in a continuous elastic medium. Phys.
Rev. E, 50, 1000-1004.

\vspace{3pt} \noindent Ruina, A.L., 1983. Slip instability and state
variable friction laws. J. Geophys. Res., 88, 10359-10370.

\vspace{3pt} \noindent Samudrala, O., Huang Y., Rosakis, A.J., 2002.
Subsonic and intersonic shear rupture of weak planes with a velocity
weakening cohesive zone. J. Geophys. Res., 107, B8 article no. 2170.

\vspace{3pt} \noindent Schallamach, A., 1971.  How does rubber
slide? Wear, 17, 301-312.

\vspace{3pt} \noindent Tsai, K.-H., Kim, K.-S., 1996. The micromechanics
of fiber pull-out. J. Mech. Phys. Solids, 44, 1147-1177.

\vspace{3pt} \noindent Tullis, T.E., 1996. Rock friction and its
implications for earthquake prediction examined via models of Parkfield
earthquakes. Proc. Natl. Acad. Sci., 93, 3803-3810.

\vspace{3pt} \noindent Xia, K., Rosakis, A.J., Kanamori, H., 2004.
Laboratory earthquakes: The sub-Rayleigh-to-supershear rupture transition.
Science, 303, 1859-1861.

\vspace{3pt} \noindent Zheng, G., Rice, J.R., 1998. Conditions under which
velocity-weakening friction allows a self-healing versus a crack-like mode
of rupture, Bull. Seismol. Soc. America, 88, 1466-1483.
}

\clearpage

\section*{Appendix: Time integration of the friction law}

The equations are integrated in time using of a rate tangent method (Peirce
et al., 1984) and automatic time step control. The sliding velocity,
$\Delta \dot{u}_{\mathrm{slip}}$, in 
the interval $[t,t+dt]$, is expressed as a linear combination of its values
at $t$ and $t+dt$ via 
\begin{equation}
\Delta \dot{u}_{\mathrm{slip}}=(1-\gamma)\Delta \dot{u}_{\mathrm{slip}}(t) +
\gamma \Delta \dot{u}_{\mathrm{slip}}(t+dt)  \label{eqtime4a}
\end{equation}
where $\gamma$ is a chosen integration parameter that can range in value
from $0$ to $1$.

A first order expansion of $\Delta \dot{u}_{\mathrm{slip}}$ 
in terms of $T_s$, $\theta_1, \theta_2$, and $\theta_0$ gives 
\begin{equation}
\Delta \dot{u}_{\mathrm{slip}}(t+dt)=\Delta \dot{u}_{\mathrm{slip}}(t)+ dt 
\Biggl( {\frac{\partial \Delta \dot{u}_{\mathrm{slip}}}{\partial{|T_s|}}} 
\mbox{{\rm sgn}}(T_s) \dot{T}_s +{\frac{\partial \Delta 
\dot{u}_{\mathrm{slip}}}{\partial{\theta_0}}} \dot{\theta}_0
+{\frac{\partial \Delta \dot{u}_{\mathrm{slip}}}{\partial{\theta_1}}}
\dot{\theta}_1 +{\frac{\partial \Delta  
\dot{u}_{\mathrm{slip}}}{\partial{\theta_2}}} \dot{\theta}_2 \Biggr )
\label{eqtime5a}
\end{equation}
where all derivatives are evaluated at $t$.

Combining (\ref{eqtime5a}) with (\ref{eqtime1}) gives 
\begin{equation}
\dot T_s=C^{\rm tan}_s \Delta \dot{u}_{s} - \dot R_p,  
\label{eqtime6a}
\end{equation}
with 
\begin{equation}
C_s^{\rm tan} = \frac{C_s}{1+\gamma d t {\frac{\partial 
\Delta \dot{u}_{\mathrm{slip}} }{\partial |T_s|}}C_s }
\end{equation}
and 
\begin{equation}
\dot R_p = C^{\rm tan}_s \mbox{{\rm sgn}}(T_s) \Biggl 
( \Delta \dot{u}_{\mathrm{slip}}^n + \gamma dt 
\displaystyle{\   {\frac{\partial \Delta \dot{u}_{\mathrm{slip}} }
{\partial  \theta_0}}}\dot{\theta}_0 + {\frac{\partial \Delta
    \dot{u}_{\mathrm{slip}} }{\partial \theta_1}} \dot\theta_1+
     {\frac{\partial \Delta  
\dot{u}_{\mathrm{slip}} }{\partial \theta_2}}\dot\theta_2 \Biggr )
\label{rate-tan}
\end{equation}
The partial derivatives in eq.~(\ref{rate-tan}) are 
\begin{eqnarray}
{\frac{\partial \Delta \dot{u}_{\mathrm{slip}} }{\partial |T_s|}} &=& 
\frac{m {V}_0\beta^{m-1}}{-T_n g(\theta_0)} \\
{\frac{\partial \Delta \dot{u}_{\mathrm{slip}} }{\partial \theta_0}} &=& 
\frac{-m {V}_0\beta^m}{g(\theta_0)} {\frac{\partial g }{\partial \theta_0}}
\\
{\frac{\partial \Delta \dot{u}_{\mathrm{slip}} }{\partial \theta_1}}= 
{\frac{\partial \Delta \dot{u}_{\mathrm{slip}} }{\partial \theta_2}}
&=& -\frac{m {V}_0\beta^m}{(\theta_1+\theta_2)}  \label{eqtime8}
\end{eqnarray}
where $g(\theta_0)$ and $\beta$ are given by eqs.~(\ref{pn2}) and
(\ref{eqbeta}), and 
\begin{equation}
\frac{\partial g}{\partial\theta_0}=\frac{ (\mu_s-\mu_d)\displaystyle 
{\frac{p}{\theta_0}} \left(\frac{L_0}{V_1\theta_0}\right)^p \exp
\left[ -\left(  
\frac{L_0}{V_1\theta_0}\right)^p\right]}{\mbox{Q}^{1/m}} + \mu^*\frac{L_0}{m
V_0 \theta_0^2}\frac{1}{\mbox{Q}^{1/m+1}}
\end{equation}
Here, 
\begin{eqnarray}
\mbox{Q}&=& \frac{L_0}{V_0 \theta} + 1 \\
\mu^* &=&\mu_d+(\mu_s-\mu_d)\exp \left[ -\left( 
\frac{L_0}{V_1\theta}\right)^p\right]
\end{eqnarray}

The integral along the interface in eq.~(\ref{eqfe2}) is written as the sum
over linear displacement interface elements. All field quantities and
constitutive state variables are known at time $t$. Also, 
the values of $\Delta \dot{u}_n$ and $\Delta \dot{u}_s$ have been
determined from solving 
eq.~(\ref{eqfe4}). Thus, $\Delta u_n$ and $\Delta u_s$ are know at time $t+dt
$. Then, at a given integration point along the interface:

\begin{enumerate}
\item  Calculate $\Delta \dot{u}_s = [\Delta u_s(t+dt)-\Delta u_s(t)]/d t$ 
and $\Delta \dot{u}_n= [\Delta u_n(t+dt)-\Delta u_n(t)]/dt$.

\item  If $\Delta u_n(t+d t) > \Sigma_0/C_n$ then set $T_n=T_s=0$ and go to
the next integration point.

\item  Using the values for the state variables at time $t$, calculate 
$g(\theta_0)$.

\item  Compute 
\begin{equation}
\beta=\frac{|T_s|}{g(\theta_0)(\theta_1+\theta_2)}.
\end{equation}

\item  If $\beta<1$ then $\Delta \dot{u}_{\mathrm{slip}}=0$. Go to \ref
{update}.

\item  If $\beta>1$ calculate $C^{tan}_s$ and $\dot{R}_p$ in eq.~(\ref
{eqtime6a}).

\item  Update the tractions and state variables via \label{update} 
\begin{eqnarray}
T_n^{k+1} &=& T_n^{k} + \dot{T}_n d t \\
T_s^{k+1} &=& T_s^{k} + \dot{T}_s d t \\
\theta_i^{k+1} &=& \theta_i^{k} + \dot{\theta}_i d t, \hspace{1mm} i=0,1,2
\end{eqnarray}
\end{enumerate}

\clearpage

\begin{table}[tbp]
\begin{center}
\begin{tabular}{|l|l|}
\hline
Property & Value \\ \hline
$\mu_s$ & $0.6$ \\ 
$\mu_d$ & $0.5$ \\ 
$V_0$ (m/s) & $100$ \\ 
$V_1$ (m/s) & $26$ \\ 
$B$ & $4.60 $ \\ 
$p$ & $1.2 $ \\ 
$m$ & $5 $ \\ 
$L_0$ (m) & $20\times10^{-6}$ \\ 
$L_1$ (m) & $20\times10^{-6}$ \\ 
$L_2$ (m) & $20\times10^{-6}$ \\ 
$C$ (MPa$^{-1}$) & $0.7$ \\ 
$D$ (MPa$^{-1}$) & $0.3 $ \\ \hline
\end{tabular}
\end{center}
\caption{Friction parameter values used in the finite element calculations.}
\label{tab:3}
\end{table}

\pagestyle{empty}

\clearpage

\thispagestyle{empty}

\clearpage
\thispagestyle{empty}

\clearpage

\clearpage

\clearpage

\clearpage

\clearpage

\clearpage

\clearpage

\clearpage

\clearpage

\clearpage

\clearpage

\clearpage
\thispagestyle{empty}

\clearpage
\thispagestyle{empty}

\clearpage

\clearpage
\thispagestyle{empty}

\clearpage
\thispagestyle{empty}

\clearpage

\clearpage

\clearpage

\end{document}